\documentclass{article} % For LaTeX2e
\usepackage{iclr2025_conference,times}

\usepackage{amsmath,amsfonts,bm}

\def\eqref#1{equation~\ref{#1}}
\def\1{\bm{1}}

\DeclareMathAlphabet{\mathsfit}{\encodingdefault}{\sfdefault}{m}{sl}
\SetMathAlphabet{\mathsfit}{bold}{\encodingdefault}{\sfdefault}{bx}{n}

\usepackage{hyperref}
\usepackage{url}

\usepackage{graphicx}
\usepackage{multirow}
\usepackage{booktabs} 
\usepackage{wrapfig}
\usepackage{subcaption}
\usepackage{pifont}

\title{Benchmarking End-To-End Performance of AI-Based Chip Placement Algorithms}

\author{
Zhihai~Wang\textsuperscript{1}$\,\,$\thanks{Equal contributions. This work was done when Zhihai Wang and Zijie Geng were interns at Huawei.},
Zijie~Geng\textsuperscript{1}$\,\,$\footnotemark[1],
Zhaojie~Tu\textsuperscript{1}$\,\,$\footnotemark[1],
Jie~Wang\textsuperscript{1}\thanks{Corresponding Author. E-mail: \texttt{jiewangx@ustc.edu.cn}},
Yuxi~Qian\textsuperscript{1},
Zhexuan~Xu\textsuperscript{1}, \\
\textbf{
Ziyan~Liu\textsuperscript{1},
Siyuan~Xu\textsuperscript{2},
Zhentao~Tang\textsuperscript{2},
Shixiong~Kai\textsuperscript{2},
Mingxuan~Yuan\textsuperscript{2},
} \\
\textbf{
Jianye~Hao\textsuperscript{2,3},
Bin~Li\textsuperscript{1},
Yongdong~Zhang\textsuperscript{1},
Feng~Wu\textsuperscript{1}
} \\
\textsuperscript{1} University of Science and Technology of China \quad 
\textsuperscript{2} Noah’s Ark Lab, Huawei \\
\textsuperscript{3} Tianjin University
}

\iclrfinalcopy % Uncomment for camera-ready version, but NOT for submission.
\begin{document}

\maketitle

\begin{abstract}
   The increasing complexity of modern very-large-scale integration (VLSI) design highlights the significance of Electronic Design Automation (EDA) technologies.
   Chip placement is a critical step in the EDA workflow, which positions chip modules on the canvas with the goal of optimizing performance, power, and area (PPA) metrics of final chip designs.
   Recent advances have demonstrated the great potential of AI-based algorithms in enhancing chip placement.
   However, due to the lengthy workflow of chip design, the evaluations of these algorithms often focus on \textit{intermediate surrogate metrics}, which are easy to compute but frequently reveal a substantial misalignment with the \textit{end-to-end performance} (i.e., the final design PPA).
   To address this challenge, we introduce ChiPBench, which can effectively facilitate research in chip placement within the AI community. ChiPBench is a comprehensive benchmark specifically designed to evaluate the effectiveness of existing AI-based chip placement algorithms in improving final design PPA metrics. Specifically, we have gathered $20$ circuits from various domains (e.g., CPU, GPU, and microcontrollers).
   These designs are compiled by executing the workflow from the verilog source code, which preserves necessary physical implementation kits, enabling evaluations for the placement algorithms on their impacts on the final design PPA.
   We executed six state-of-the-art AI-based chip placement algorithms on these designs and \emph{plugged} the results of each single-point algorithm into the physical implementation workflow to obtain the final PPA results.
   Experimental results show that even if intermediate metric of a single-point algorithm is dominant, while the final PPA results are unsatisfactory.
   This suggests that the AI community should concentrate more on enhancing end-to-end performance rather than those intermediate surrogates.
   We believe that our benchmark will serve as an effective evaluation framework to bridge the gap between academia and industry.
   \end{abstract}

   \section{Introduction}
\begin{figure}[t]
       \centering
       \includegraphics[width=\textwidth]{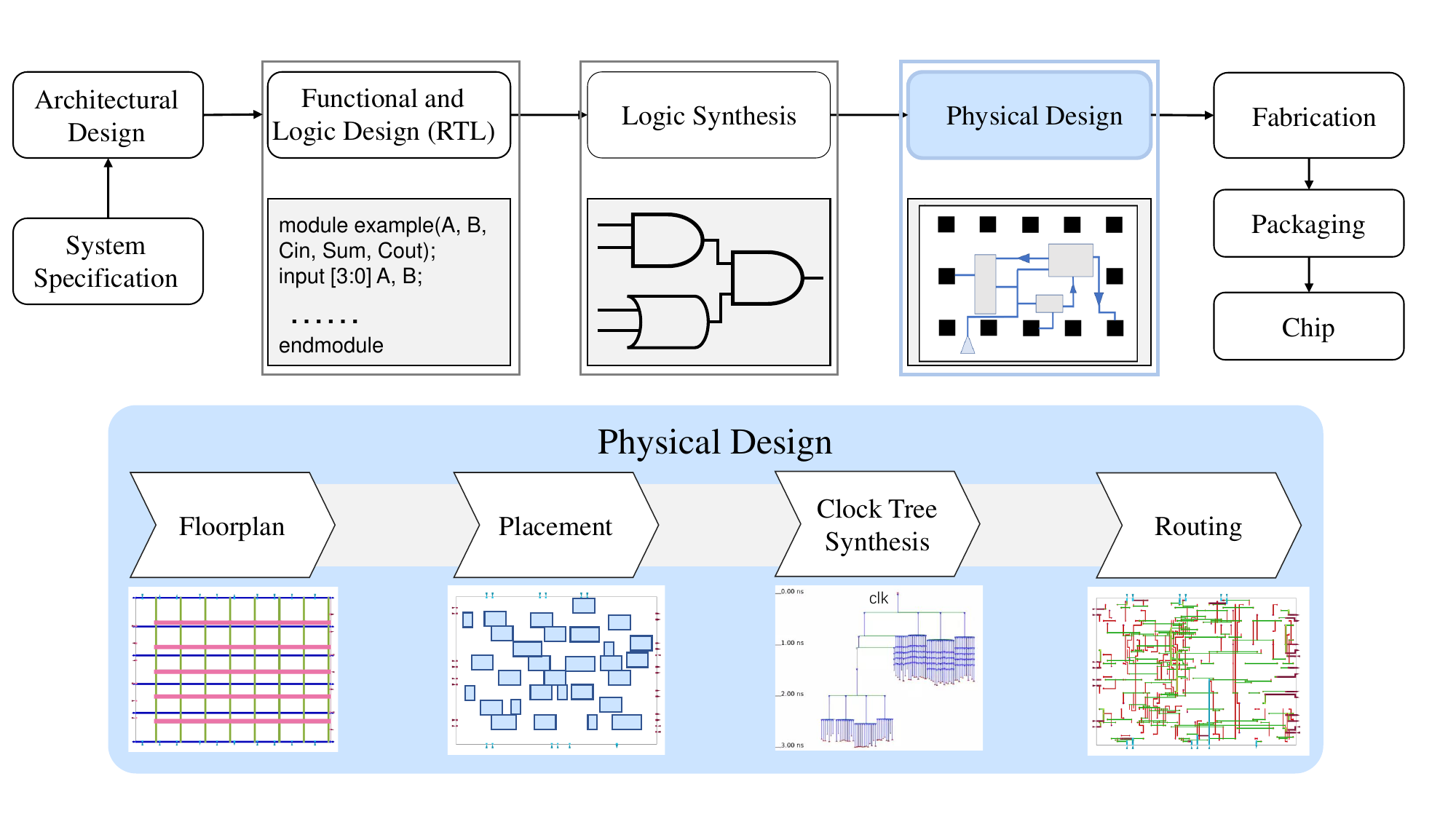}
       \caption{Illustration of the modern chip design workflow.}
       \label{fig:flowchart}
   \end{figure}

   The exponential growth in the scale of integrated circuits (ICs), in accordance with Moore’s law, has posed significant challenges to chip design~\citep{huang2021machine, lopera2021survey}.
   To handle the increasing complexity, many electronic design automation (EDA) tools have been developed to assist hardware engineers.
   As shown in Figure~\ref{fig:flowchart}, EDA tools automate various steps in the chip design workflow, including high-level synthesis, logic synthesis, physical design, testing and verification \citep{huang2021machine, sanchez2023comprehensive}.
   
   Chip placement is a critical step in the chip design workflow, which aims to position chip modules on the canvas, with the goal of optimizing the performance, power, and area (PPA) metrics of final chip designs~\citep{cheng2023assessment,bbo2023,dreamplace2019}.
   Traditionally, this is done manually by human expert designers, which costs much labor and necessitates much expert prior knowledge.
   Therefore, a lot of design automation methods, especially those AI-based algorithms, have been developed to automate this process.
   These methods mainly fall into two categories: optimization-based methods and reinforcement learning (RL)-based methods~\citep{geng2024reinforcement}.
   Optimization-based methods employ traditional optimization algorithms, such as simulated annealing (SA)~\citep{rlea2020} and evolutionary algorithms (EA)~\citep{bbo2023} to directly address the large-scale optimization problem, exploring the design space to identify near-optimal solutions.
   In recent research, macro placement has been formulated as a Markov Decision Process (MDP), where the macro positions are determined sequentially~\citep{graphplace2021}.
   Reinforcement learning (RL) has emerged as a promising technique for this task due to its ability to continuously improve performance based on feedback from the environment through trial and error~\citep{deeppr2021,prnet2022,maskplace2022,chipformer2023}.
   
   However, due to the lengthy workflow of chip design, the evaluations of these algorithms often focus on \textit{intermediate surrogate metrics}, which are easy to compute but frequently reveal a substantial misalignment with the \textit{end-to-end performance} (i.e., the final design PPA).
   On one hand, obtaining the end-to-end performance of a given chip placement solution requires a large amount of engineering efforts due to the lengthy workflow of chip design. In particular, we found that directly using the existing open-source EDA tools and chip placement datasets even usually fails to obtain the end-to-end performance. Thus, existing AI-based chip placement algorithms~\citep{chipformer2023,geng2024reinforcement} train and evaluate learned models using intermediate surrogate metrics which are simple and easy to obtain.
   On the other hand, as the PPA metrics are reflected by many aspects that have not been adequately considered in the previous stages, there exhibits a critical gap between the surrogate metrics and the final PPA objectives. Therefore, this gap significantly limits the use of existing AI-based placement algorithms in practical industrial scenarios.
   
   To address this challenge, we propose ChiPBench, a comprehensive benchmark specifically designed to evaluate the effectiveness of existing AI-based chip placement algorithms in improving final design PPA metrics.
   Appealing features of ChiPBench include its fully open source and reproducible characteristics, covering the entire EDA workflow from the source verilog code, and unifying the evaluation framework of AI-based chip placement methods using end-to-end performance.
   Thus, ChiPBench can effectively facilitate research in chip placement within the AI community by taking the first step towards a fully reproducible unified evaluation framework using the end-to-end performance. In terms of the dataset, we have gathered $20$ circuits from various domains (e.g., CPU, GPU, and microcontrollers).
   Then, these designs are compiled by executing the workflow from the verilog source code, which preserves sufficient physical implementation kits, enabling evaluations for the placement algorithms on their impacts on the final design PPA.
   In terms of the evaluated algorithms, we executed \textit{six} state-of-the-art AI-based chip placement algorithms on the aforementioned designs and \emph{plugged} the results of each single-point algorithm into the physical implementation workflow to obtain the final PPA results.
   
   Experimental results show that even if intermediate metric of a single-point algorithm is dominant, while the final PPA results are unsatisfactory. Moreover, visualization experiments demonstrate that intermediate metrics have weak correlation with the final design PPA, emphasizing the importance of developing algorithms towards optimizing the final design PPA rather than intermediate metrics. This suggests that the AI community should concentrate more on enhancing end-to-end performance rather than those intermediate metrics.
   We believe that our benchmark will serve as an effective evaluation framework to bridge the gap between academia and industry.
   
   We summarize our major contributions as follows. 
   (1) Our proposed ChiPBench is a reproducible and unified evaluation framework of existing AI-based chip placement algorithms using end-to-end performance with fully open source EDA tools. This can effectively facilitate research in chip placement within the AI community. 
   (2) We collected $20$ circuits from various domains, and construct a dataset by executing the EDA workflow from the verilog source code, preserving sufficient physical implementation kits for end-to-end performance evaluation. 
   (3) We evaluate six state-of-the-art AI-based chip placement algorithms, including most popular AI-based chip placement algorithms.
   (4) Experiments demonstrate that intermediate metrics have weak correlation with the final design PPA, emphasizing the critical importance of developing algorithms towards optimizing the final design PPA rather than intermediate metrics.
   
   \section{Related Work}
   \textbf{Datasets}
   Some well-known EDA conferences, such as ISPD and ICCAD, host contests addressing EDA challenges and offer benchmarks with processed data for researchers.
   However, in the early years (e.g., ISPD2005~\citep{ispd2005} and ICCAD2004~\citep{adya2009iccad}), the provided datasets used overly simplified \texttt{Bookshelf} formats, which are abstracted versions of the actual design kits.
   Therefore, we cannot evaluate the final PPA of the placement results on those datasets. 
   Recently, ISPD2015~\citep{2015ispd} and ICCAD2015~\citep{iccad2015} have offered benchmarks and datasets closer to real-world applications, including necessary netlist, library, and design exchange files, broadening their utility slightly.
   Nevertheless, they still lack the essential information (e.g., necessary design kits) to run the open-source EDA tools such as OpenROAD~\citep{kahng2021openroad}. 
   Beyond these conferences, some other datasets have been developed in various directions.
   For example, the EPFL~\citep{epfl} benchmarks and the larger OpenABC-D \cite{openabc} dataset concentrated on synthetic netlists, primarily for testing modern logic optimization tools with a focus on logic synthesis.
   CircuitNet 2.0 \citep{circuitnet}, on the other hand, shifted the focus towards providing multi-modal data for prediction tasks, enhancing the capability for various prediction tasks through the use of diverse data modalities.
Compared with previous efforts, our proposed dataset focuses on the entire EDA workflow. As shown in Table~\ref{comparison}, a comparison between our dataset and mainstream datasets highlights the distinctions. It provides complete files for each case and necessary design kits, such as timing constraints, library files, and LEF files, offering a comprehensive dataset that supports all stages of physical implementation and fosters a more integrated approach to chip design and evaluation.

% \usepackage{booktabs}
% \usepackage{booktabs}
% \usepackage{booktabs}
\begin{table}[t]
 \caption{Comparison Between Our Dataset and Existing Datasets}
       \resizebox{0.99\textwidth}{!}{
\begin{tabular}{@{}cccccc@{}}
\toprule
\textbf{Dataset}          & \begin{tabular}[c]{@{}c@{}}Complete \\ Design Suite\end{tabular} & \begin{tabular}[c]{@{}c@{}}Logic Synthesis \\ Support\end{tabular} & \begin{tabular}[c]{@{}c@{}}Physical Design \\ Support\end{tabular} & \begin{tabular}[c]{@{}c@{}}Full EDA Flow \\ Support\end{tabular} & \begin{tabular}[c]{@{}c@{}}Large Scale \\ \&Diversity\end{tabular} \\ \midrule
\textbf{ISPD2005}         & \ding{55}                                                        & \ding{55}                                                          & \ding{51}                                                          & \ding{55}                                                        & \ding{55}                                                         \\
\textbf{ICCAD2015}        & \ding{55}                                                        & \ding{55}                                                          & \ding{51}                                                          & \ding{55}                                                        & \ding{51}                                                         \\
\textbf{EPFL Benchmarks}  & \ding{55}                                                        & \ding{51}                                                          & \ding{55}                                                          & \ding{55}                                                        & \ding{55}                                                         \\
\textbf{OpenABC-D}        & \ding{55}                                                        & \ding{51}                                                          & \ding{55}                                                          & \ding{55}                                                        & \ding{51}                                                         \\
\textbf{CircuitNet 2.0}   & \ding{55}                                                        & \ding{55}                                                          & \ding{51}                                                          & \ding{55}                                                        & \ding{51}                                                         \\
\textbf{ChiPBench (Ours)} & \ding{51}                                                        & \ding{51}                                                          & \ding{51}                                                          & \ding{51}                                                        & \ding{51}                                                         \\ \bottomrule
\end{tabular}}
\label{comparison}
\end{table}

   \textbf{Placement Algorithms}
   Recent advancements in AI technology within the EDA field have led to a variety of AI-based chip placement algorithms. 
   (1) Black-Box Optimization methods.
   Simulated Annealing~\citep{cheng2023assessment} provides a probabilistic method for finding a good approximation of the global optimum.
   Wire-Mask-Guided Black-Box Optimization~\citep{bbo2023} uses a wire-mask-guided greedy procedure to optimize macro placement efficiently.
   (2) Analytical methods.
   DREAMPlace~\citep{dreamplace2019} uses deep learning toolkits to achieve over a 30x speedup in placement tasks. 
   AutoDMP~\citep{autodmp2023} leverages DREAMPlace for the concurrent placement of macros and standard cells, enhancing macro placement quality.
   (3) Reinforcement Learning methods.
   MaskPlace~\citep{maskplace2022} treats chip placement as a visual representation learning problem, reducing wirelength and ensuring zero overlaps.
   ChiPFormer~\citep{chipformer2023} employs offline reinforcement learning, fine-tuning on unseen chips for better efficiency.
   The evaluation of these algorithms mainly focuses on intermediate metrics.
   In contrast, we utilized the \textit{end-to-end performance} to evaluate six existing AI-based chip placement algorithms, encompassing a significant portion of mainstream AI-based placement algorithms.

   \section{Background on Electronic Design Automation}
   Electronic Design Automation (EDA) is a suite of software tools vital for designing and developing electronic systems, primarily integrated circuits (ICs).
   These tools enable electrical engineers to efficiently transform innovative concepts into functional products, addressing the increasing complexity and demands of modern chip design.
   EDA optimizes the entire design process from schematic capture to layout and fabrication, reducing time-to-market and enhancing design precision and sophistication.
   In the chip design workflow, EDA tools support various functions: they perform simulations to verify circuit behavior, execute synthesis to convert high-level descriptions to gate-level implementations, and manage physical layouts to ensure designs can be realized in silicon.
   % These capabilities are crucial as they streamline the transition from conceptual designs to physical chips, encompassing stages from the Register Transfer Level (RTL) to the final GDSII file.
   % The RTL uses Hardware Description Languages (HDLs) like Verilog or VHDL to define hardware functions, simplifying the design by abstracting complex, transistor-level details and focusing on architecture. The GDSII file format, crucial for manufacturing, encodes the complete design in a binary format sent to foundries, ensuring that the final chip conforms to its design specifications.
   As shown in Figure~\ref{fig:flowchart}, the EDA design flow includes several key stages: logic synthesis, floorplanning, placement, Clock Tree Synthesis (CTS), and routing.
   Below are concise descriptions of each stage, illustrating their importance in the integrated circuit development process.
   
   \textbf{Logic Synthesis} transforms a high-level circuit description into an optimized gate-level netlist~\citep{berndt2022review,wang2024a_prunex, wang2024a_have}.
\textbf{Floorplan} involves deciding the layout of major components within an integrated circuit, positioning blocks and core components to balance signal integrity, power distribution, and area utilization. 
   % A well-designed floorplan helps in reducing wire lengths, which can improve performance and reduce power consumption and delays. It also sets the framework for more detailed placement and routing stages.
   \textbf{Placement} involves assigning specific locations to various circuit components within the core area of the chip, following the floorplanning stage. The primary objective of this stage is to strategically place the components to optimize performance metrics such as delay and power consumption while ensuring adherence to design rules~\citep{geng2024reinforcement}.
   \textbf{Clock Tree Synthesis (CTS)} creates a clock distribution network within an IC to minimize those clock effects, and ensure the correct timing synchronization for circuit operation.
   \textbf{Routing} involves creating the physical paths for electrical connectivity between various components on the IC as per the netlist. This stage must handle multiple layers of the chip, avoid obstacles, manage signal integrity, and meet all electrical and timing constraints~\citep{prnet2022}. 

%    \textbf{Logic Synthesis} transforms a high-level circuit description into an optimized gate-level netlist~\citep{berndt2022review,wang2024a_prunex, wang2024a_have}.
% \textbf{Floorplan} involves deciding the layout of major components within an integrated circuit, positioning blocks and core components to balance signal integrity, power distribution, and area utilization.
% \textbf{Placement} assigns specific locations to circuit components within the core area of the chip, aiming to optimize performance metrics like delay and power consumption while following design rules~\citep{geng2024reinforcement}.
% \textbf{Clock Tree Synthesis (CTS)} creates a clock distribution network to minimize clock effects and ensure timing synchronization.
% \textbf{Routing} creates the physical paths for electrical connectivity between components as per the netlist, managing multiple layers, avoiding obstacles, and meeting electrical and timing constraints~\citep{prnet2022}.
   % Routing is divided into global routing and detailed routing. Global routing performs coarse-grained routing by dividing the entire canvas into multiple GCells and routing only specific networks between these GCells. Detailed routing finishes the remaining nets and may attempt to re-route the nets that are more critical to optimize performance and meet design constraints.
   % Previous process regards the clock as ideal which ignores skew and jitter, thus the CTS 
   % Effective placement is critical as it directly affects the performance of the chip and the ease of routing.
   
   \textbf{Chip Placement} The placement stage is crucially divided into two distinct phases: \textit{macro placement} and \textit{cell placement}.
   (1) Macro placement is a critical very large-scale integration (VLSI) physical design problem that targets the arrangement of larger components, such as SRAMs and clock generators—often called macros. This phase significantly impacts the chip's overall floorplan and essential design parameters like wirelength, power, and area. 
   (2) Following this, the standard cell placement phase addresses the arrangement of the more numerous and smaller standard cells, which serve as the fundamental building blocks of digital designs. This phase typically utilizes analytical solvers to secure an optimized configuration that not only minimizes wirelength but also enhances the electrical and timing performance of the chip.

   \section{Dataset}
   \subsection{Description of Designs}
   Due to the oversimplification of datasets in early years, there exists a significant gap between these datasets and real-world applications.
   For instance, the usually used \texttt{Bookshelf} format~\citep{ispd2005,adya2009iccad} is overly simplified so that placement results given in such format are inapplicable for the subsequent stages to obtain a valid final design.
   Some later datasets~\citep{iccad2015} provide the \texttt{LEF/DEF} and necessary files for running these stages, but the contained circuits are still limited and they still lack some information for open-source tools like OpenROAD to work. For instance, the library file lacks buffer definitions, which is necessary for the clock tree synthesis phase, and the LEF file has incomplete layer definitions, which hinders the routing phase. To address this issue, we construct a dataset with comprehensive physical implementation information across the entire flow.
   Our dataset involves collecting a series of designs spanning various domains, including components such as CPUs, GPUs, network interfaces, image processing technologies, IoT devices, cryptographic units, and microcontrollers.
   Additionally, the dataset features a diverse array of sizes, with cell ranges from thousands to nearly a million.
   Our dataset features a complete design suite that supports the full EDA flow and includes a diverse range of sizes and domains, as illustrated in Table~\ref{comparison}.
   The statistics for each case is detailed in Table~\ref{table1}, and we defer more details to Table~\ref{table2} in Appendix~\ref{app_more_results}.
   
   %iccad15:
   %- lib库中缺失buffer定义,导致无法进行cts
   %- lef文件中layer定义不完整,无法进行routing
   %- openroad中不支持iccad15的sdc的语法(这点可能可以不写上去?)
   
   %our dataset包括:
   %- lib库中有完整的元件的物理信息
   %- lef有完整floorplan的信息,有layer的定义:metal,via
   %- 有sdc文件,时序信息

   \begin{table}[t]
       \centering
       \caption{Statistics of designs used in our benchmark.}
       \resizebox{0.99\textwidth}{!}{
       \begin{tabular}{@{}cccccccc@{}}
       \toprule
       \toprule
           Id & Design & \#Cells & \#Nets & \#Macros & \#Pins & \#IOs & \#Edges \\ \hline
           1 & 8051~\citep{ref6} & 13865 & 16424 & 0 & 50848 & 10 & 16174 \\ 
           2 & ariane136~\citep{ref11}& 175248 & 191081 & 136 & 609834 & 495 & 187911 \\ 
           3 & ariane133~\citep{ref11}& 168551 & 184856 & 132 & 592261 & 495 & 183142 \\ 
           4 & bp~\citep{ref11}& 301030 & 333364 & 24 & 984093 & 1198 & 333364 \\ 
           5 & bp\_be~\citep{ref11}& 50881 & 58428 & 10 & 182949 & 3029 & 58092 \\ 
           6 & bp\_fe~\citep{ref11}& 33206 & 36379 & 11 & 111510 & 2511 & 36203 \\ 
           7 & CAN-Bus~\citep{ref1} & 815 & 935 & 0 & 2637 & 13 & 935 \\ 
           8 & DE2\_CCD\_edge~\citep{ref8}& 2333 & 3270 & 0 & 7823 & 64 & 3170 \\ 
           9 & dft48~\cite{dft}  & 48488 & 52575 & 68     & 125501 & 132 & 50654 \\
           10 & FPGA-CAN~\citep{ref3} & 140848 & 178913 & 0 & 532024 & 4 & 176472 \\ 
           11 & iot shield~\citep{ref2} & 904 & 1006 & 0 & 2995 & 33 & 974 \\ 
           12 & mor1kx~\citep{ref12}&104293&130743&0&374983&576&125979 \\ 
           13 & or1200~\cite{or1200} & 43386 & 32195 & 20     & 97047  & 383 & 31958 \\
           14 & OV7670\_i2c~\citep{ref7}& 332 & 340 & 0 & 979 & 29 & 316 \\ 
           15 & picorv~\citep{ref9}& 8851 & 10531 & 0 & 32195 & 409 & 10470 \\ 
           16 & serv~\citep{ref10}& 1291 & 1482 & 0 & 3915 & 306 & 1403 \\ 
           17 & sha256~\citep{ref5} & 10120 & 12283 & 0 & 38758 & 77 & 12176 \\ 
           18 & subrisc~\citep{ref4} & 859382 & 1103295 & 0 & 3359066 & 34 & 1092653 \\ 
           19 & swerv\_wrapper~\citep{ref11}& 96435 & 105026 & 28 & 354652 & 1416 & 104565 \\ 
           20 & toygpu~\citep{ref13}&368081&466513&0&1399167&11&461675 \\ 
           
           \bottomrule
       \end{tabular}}
       \label{table1}
   \end{table}

   % \begin{table}[t]
   % \centering
   % \label{tab:main_results}
   % \caption{Main Results}
   % \resizebox{0.99\textwidth}{!}{
   % \begin{tabular}{@{}cccccccccc@{}}
   % \toprule
   % \toprule
   % \textbf{Method} & \textbf{MacroHPWL} $\downarrow$ & \textbf{HPWL} $\downarrow$ & \textbf{Wirelength} $\downarrow$ & \textbf{Congestion} $\downarrow$ & \textbf{Power} $\downarrow$ & \textbf{WNS} $\downarrow$ & \textbf{TNS} $\downarrow$ & \textbf{NVP} $\downarrow$ & \textbf{Area} $\downarrow$ \\ \midrule
   % \textbf{ChiPFormer} & 0.918 & 1.045 & 1.073 & 1.073 & 1.045 & 1.280 & 1.437 & 0.995 & 1.009 \\
   % \textbf{BBO} & 0.944 & 1.042 & 1.387 & 1.358 & 0.928 & 2.093 & 0.957 & 0.739 & 1.008 \\
   % \textbf{SA} & 1.107 & 1.145 & 1.258 & 1.267 & 1.145 & 1.136 & 1.168 & 0.864 & 1.008 \\
   % \textbf{DREAMPlace} & 1.378 & 1.020 & 1.125 & 1.130 & 1.055 & 1.300 & 2.157 & 1.183 & 1.008 \\
   % \textbf{MaskPlace} & 2.395 & 1.077 & 1.084 & 1.084 & 1.038 & 1.196 & 1.410 & 1.028 & 1.005 \\ \midrule
   % \textbf{OpenROAD} & 1.000 & 1.000 & 1.000 & 1.000 & 1.000 & 1.000 & 1.000 & 1.000 & 1.000 \\ \bottomrule
   % \end{tabular}
   % }
   % \end{table}

   \subsection{Dataset Generation Pipeline}
   \label{sec:pipeline}
   We use OpenROAD~\citep{kahng2021openroad}, an open-source EDA tool, for generating our dataset.
   OpenROAD integrates various tools, such as Yosys~\citep{wolf2016yosys} for logic synthesis, TritonMacroPlacer for macro placement, RePlAce~\citep{replace2018} for cell placement, TritonCTS for clock tree synthesis, and TritonRoute for detailed routing.
   The choice of open-source tool allows for full reproducibility of our results and supports the promotion of the open-source community, ensuring that all generated data and methodologies are open-source.
   The initial dataset generation starts with Verilog files as raw data.
   OpenROAD performs logical synthesis to convert these high-level descriptions into a netlist, detailing the electrical connections among circuit components.
   This netlist is then used by OpenROAD's integrated floorplanning tool to configure the physical layout of the circuit on silicon.
   The resulting design from the floorplanning stage is converted into \texttt{LEF/DEF} files by OpenROAD, facilitating the application of subsequent placement algorithms.
   Simultaneously, we complete the EDA design flow through OpenROAD, generating data at subsequent stages, including placement, CTS, and routing.
   
   \section{Algorithms}
   AI-based chip placement algorithms can be roughly grouped into three categories: black-box optimization (BBO) methods, analytical methods (gradient-based methods), and reinforcement learning (RL) methods. Each category frames the placement task as an optimization problem but adopts distinct objectives and methodologies. We present details as follows.
   
   \subsection{Black-Box-Optimization (BBO) Methods}
   A straightforward intuition is to view the chip placement task as a black-box-optimization (BBO) problem, where the inner workings of the objective functions are inaccessible, and solutions are evaluated only based on the output metrics.
   
   \textbf{Simulated Annealing (SA)} is a heuristic BBO optimization algorithm favored for its simplicity in implementation.
   Specifically, the SA algorithm generates solutions by perturbing the solution space and then assessing the resulting representation.
   Different methods have been developed to effectively map representations to placement solutions~\citep{kirkpatrick1983optimization,sherwani2012algorithms,ho2004orthogonal,shunmugathammal2020novel,rlea2020}, such as sequence pair~\citep{sequencepair1996} and B$^\ast$-tree~\citep{btree2000}. 
   Solutions are probabilistically accepted based on an annealing temperature to escape local optima in pursuit of a global optimum.
   Due to its simplicity in implementation, the SA algorithm often serves as a strong baseline in previous studies.
   In this work, we incorporate a specific SA implementation~\citep{cheng2023assessment} utilizing operations like swaps, shifts, and shuffles, and a cost function that balances wirelength, density, and congestion.
   
   %Simulated Annealing (SA) is a heuristic-based global optimization algorithm. The fundamental idea is to introduce randomness into the search process to prevent the algorithm from prematurely converging to a local optimum rather than the global optimum. 
   
   %Due to its effectiveness in avoiding local optima, broad applicability, minimal parameters, and simplicity, Simulated Annealing is widely used for BBO problems. However, due to its slow convergence rate and computational inefficiency, as well as its high dependence on parameter selection, it is often used as a baseline algorithm in placement.
   
   %There are many existing SA algorithms, and the method adopted in this paper is~\cite{graphplace2021}, which is used as one of the baselines, includes actions such as swap, shift, mirror, move, and shuffle. The cost function is defined as a weighted average of wirelength, density, and congestion.

   % \textbf{WireMask-EA}~\citep{bbo2023} is a BBO framework recently appearing on the AI conference NeurIPS'2023.
   % It leverages wiremask to greedily guide the mapping from genotypes to phenotypes.
   % Here the concept of wiremask was first proposed by \citet{maskplace2022}, and it is a matrix representing the potential increases in HPWL when placing the next macro on the canvas.
   % Building on this, \citet{bbo2023} implement different kinds of BBO algorithms, including random search (RS), evolutionary algorithm (EA), and Bayesian optimization (BO).
   \textbf{WireMask-EA}~\citep{bbo2023} is a BBO framework that was recently introduced at the NeurIPS 2023 conference, positioning itself as an innovative approach in the intersection of AI and EDA. The framework utilizes a novel concept called wiremask, which plays a crucial role in guiding the mapping process from genotypes to phenotypes in a greedy manner. The wiremask concept was originally introduced by \citet{maskplace2022}, where it is defined as a matrix that predicts the potential increase in Half-Perimeter Wirelength (HPWL) for each subsequent macro placement on the design canvas. By estimating the wirelength increase, the wiremask helps in making informed decisions during the placement process, thereby potentially improving the quality of the layout. Building upon this concept, \citet{bbo2023} extended the framework to integrate several types of Black-Box Optimization (BBO) algorithms, including random search (RS), evolutionary algorithm (EA), and Bayesian optimization (BO), demonstrating the versatility of the approach in handling complex optimization tasks in chip design.

   \subsection{Analytical (Gradient-Based) Methods}
   Analytical methods formulate the optimization objective as an analytical function of module coordinates.
   This formulation enables efficient solutions through techniques like quadratic programming~\citep{aplace2005,rql2007,fastplace2007,kraftwerk2008,ntuplace2008,maple2012,complx2012} and direct gradient descent~\citep{eplace2014,replace2018,dreamplace2019,dreamplace2.0-2020,dreamplace3.0-2020,dreamplace4.0-2022}.
   This work focuses on the gradient-based algorithms, which are by far the more mainstream algorithms.
   
   \textbf{DREAMPlace}~\citep{dreamplace4.0-2022} is a GPU-accelerated framework that leverages differentiable proxies, such as approximate HPWL, as optimization objectives.
   It was built upon the previous analytical placement algorithms, ePlace~\citep{eplace2014} and RePlAce~\citep{replace2018}, yet significantly speeding up the placement process by using GPUs for acceleration.
   The series of versions of DREAMPlace introduces diverse differentiable proxies to better align the PPA improvement.
   
   \textbf{AutoDMP}~\citep{autodmp2023} extends DREAMPlace by automating hyperparameter tuning through multi-objective Bayesian optimization.
It further accelerates the optimization process and reduces manual tuning efforts. At that time, this work showcased the promising potential of integrating GPU-accelerated algorithms with machine learning techniques for automating VLSI design.
   
   \subsection{Reinforcement Learning (RL) Methods}
   As VLSI systems grow in complexity, RL methods are being explored to enhance placement quality.
   GraphPlace~\citep{graphplace2021} first models macro placement as a RL problem.
   Subsequently, DeepPR~\citep{deeppr2021} and PRNet~\citep{prnet2022} establish a streamlined pipeline encompassing macro placement, cell placement, and routing.
   However, they treat density as a soft constraint, which may violate non-overlap constraint during training.
   Therefore, in this work, we mainly focus on MaskPlace and ChiPFormer, which are recent SOTA algorithms with hard non-overlapping constraints.
   
   \textbf{MaskPlace}~\citep{maskplace2022} represents the chip states as pixel-level visual inputs, including a wiremask (recording the HPWL increment for each grid), the viewmask (a global observation of the canvas), and the positionmask (to ensure non-overlapping constraint). 
   Furthermore, it uses dense reward to boost the sample efficiency.
   
   % MaskPlace, an RL-based model, introduces a new perspective on AI-assisted chip placement. It represents the chip states as pixel-level visual inputs, which include position, wirelength, and view information. This approach helps the model to comprehensively capture all pin position information, maintain the complete action space on the expansive chip canvas, and ensure that the layout remains free of overlaps simultaneously. Additionally, it uses dense reward to boost the sample efficiency, enhancing the overall effectiveness of the model.
   
   \textbf{ChiPFormer}~\citep{chipformer2023} represents the first offline RL method.
   It is pretrained on various chips via offline RL and then fine-tuned on unseen chips for better efficiency. 
   As a result, the time for placement is significantly reduced.
   
   % ChiPFormer, integrating a transformer model for the first time in offline RL, learns a transferable placement strategy from expert human layouts. It's pretrained on a variety of chips using existing layout data, enhancing both time and sample efficiency, and fine-tuned on unseen chips to optimize performance. This approach allows for high-quality, rapid placements, significantly reducing the time required for new chip designs. ChiPFormer showcases robust Zero-Shot transfer capabilities, making it ideal for quick, efficient chip layout processes.

   \section{Evaluation}
   \subsection{Evaluation Metrics}
   \subsubsection{Final Design PPA Metrics}
   The primary goal of the entire Electronic Design Automation (EDA) workflow is to optimize the final PPA metrics.
   PPA stands for performance, power, and area—three crucial dimensions used to evaluate the quality of a chip product.
   These dimensions are assessed using several critical metrics, including worst negative slack (WNS), total negative slack (TNS), number of violating paths (NVP), power, and area.
   Optimizing these PPA metrics has been a major focus in the industry, approached through expert-designed heuristics.
   However, the challenge of PPA optimization has not been fully recognized within the AI community.
   Bridging this gap and improving the incorporation of AI strategies into PPA optimization are key goals of this benchmark.
   
   In terms of specific metrics, Worst Negative Slack (WNS) and Total Negative Slack (TNS) are essential for assessing the timing performance of a chip circuit. Slack is the discrepancy between the expected and required arrival times of a signal, with negative slack indicating a timing violation. WNS pinpoints the most severe negative slack within a circuit, thus identifying the most critical timing issue. Conversely, TNS aggregates all negative slacks, providing a comprehensive view of the circuit's overall timing challenges. Moreover, the number of violating paths (NVP) counts the paths that fail to meet the timing constraints, further illustrating the timing performance issues.
   
   \subsubsection{Intermediate Metrics}
   Commonly used intermediate surrogate metrics include Congestion,  Wire Length (WL), Half Perimeter Wire Length (HPWL), and Macro HPWL (mHPWL). 
   Congestion evaluates the density of wires in different chip regions.
   High congestion in certain areas can pose substantial challenges during the routing stage.
   While not a direct component of the PPA metrics, managing congestion effectively is essential to ensure that the chip can be successfully manufactured.
   Therefore, it is also considered as an evaluation metric in this paper.
   Congestion is typically estimated after the Clock Tree Synthesis (CTS) stages but before the detailed routing stage, allowing for adjusting macro placement and routing strategies to mitigate potential issues.
   
   Wire Length (WL) is the total length of all wires connecting all modules in a chip.
   Half Perimeter Wire Length (HPWL) is the sum of half perimeters of bounding boxes that encompass all pins in each net.
   It is widely used as an estimation of WL and is obtained after cell placement.
   Macro HPWL (mHPWL) further simplifies HPWL by only considering the macros.
   It is favored in recent studies as it can be immediately obtained after macro placement.
   These metrics are thought to correlate with the final PPA, but they do not directly reflect the chip quality.

   \subsection{End-to-End Evaluation Workflow}
      \vspace{-2mm}
   \begin{wrapfigure}{r}{0.4\textwidth}
       \centering
       \includegraphics[width=0.4\textwidth]{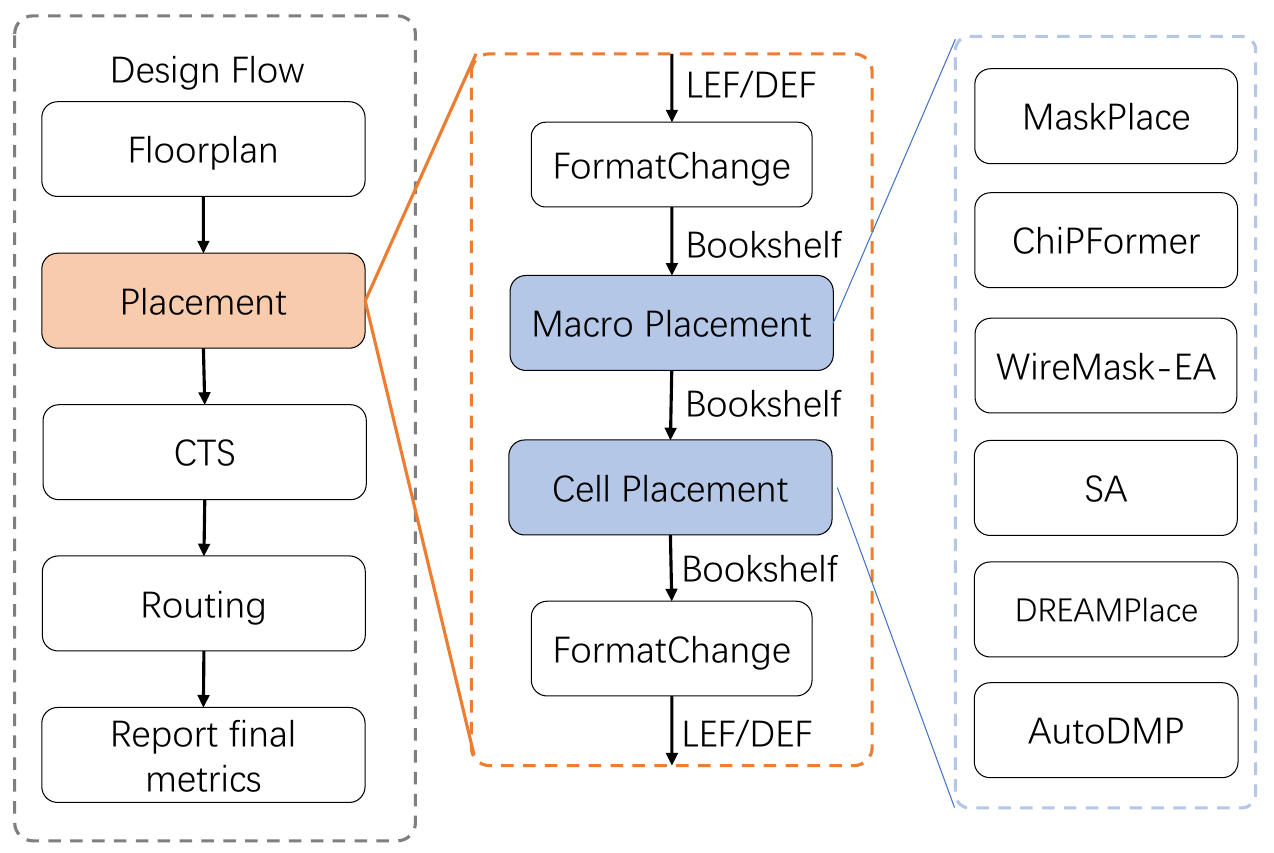}
       \vspace{-2mm}
       \caption{Illustration of our end-to-end evaluation workflow.}
       \vspace{-2mm}
       \label{fig:workflow}
   \end{wrapfigure}
   We present an end-to-end evaluation workflow utilizing OpenROAD-flow-scripts \citep{kahng2021openroad} for the various stages of the EDA design flow, as illustrated in Figure~\ref{fig:workflow}.
   All tools used in this workflow are open-source, providing a significant advantage over other workflows that rely on commercial software. This workflow is designed to offer a comprehensive assessment of optimization algorithms at any stage of the design flow.

   Our dataset comprises design kits needed for each stage of the physical design flow.
   For evaluating any stage-specific algorithm, the output file from the preceding stage serves as the input for the algorithm under evaluation. The algorithm processes this input to generate its output, which is subsequently plugged into the OpenROAD design flow.
   Ultimately, final performance metrics such as TNS, WNS, Area, and Power are reported, providing a comprehensive end-to-end performance assessment.This method offers a holistic set of metrics that can evaluate the optimization effects of any stage-specific algorithm on the final chip design, providing consistency of metrics and avoiding the limitations of overly simplified metrics confined to a single stage.It is particularly beneficial for the optimization and development of various algorithms, ensuring their improvements translate to practical enhancements in chip design and foster the development of more efficient and effective open-source EDA tools through a robust framework for testing and improvement.
   
   \subsection{Experimental Setup}
   \label{sec:exp}
   We apply the aforementioned workflow to evaluate six macro placement algorithms: SA, WireMask-EA, DREAMPlace, AutoDMP, MaskPlace, ChiPFormer,  and the default algorithm in OpenROAD. Additionally, we also assess cell placement algorithms for DREAMPlace and AutoDMP.
   % 这里提下原因：因为之前很多算法其实都是用bookshelf格式 due to the reasons.. 但是我们这里都是标准的design exchange format，所以我们将design抽象化成bookshelf for consistent in order to evaluate these single-point-algorithm。。。
   As most of these methods only support the circuit data in a \texttt{BookShelf} format, while the circuits in our used dataset are in a standard \texttt{LEF/DEF}, we start by converting the \texttt{LEF/DEF} files from the floorplan stage of our dataset to \texttt{BookShelf} format to serve as the input for the placement algorithms.
   After finishing the macro placement stage, the resulting placement files are then converted back to \texttt{DEF} format and reintroduced into the original flow. 
   The resulting placement files in \texttt{BookShelf} format are then converted back to \texttt{DEF} and reintroduced into the original flow.
   % In this stage it conduct two different operations, coarse placement and legalization, with the coarse placement splitting the cells simultaneously using gradient descent algorithms and the legalization stage correct the positions of cells subject to the non-overlapping constraint.
   Finally, we report the final metrics, obtaining end-to-end evaluation results.
   Additionally, we perform global placement and detailed placement using OpenROAD’s native Place method and complete the entire flow to obtain the final metrics for comparison with other algorithms. Our project is open-sourced on \href{https://anonymous.4open.science/r/ChiPBench-D1B2/}{GitHub}.

   \section{Results and Discussions} % Results
   \subsection{Main Results}

   \textbf{Macro Placement} We evaluate the AI-based chip placement algorithms, including  SA, WireMask-EA, DREAMPlace, AutoDMP, MaskPlace, and ChiPFormer, using both intermediate metrics and end-to-end performance.
   The results for macro placement are in Table \ref{tab_main_results_macro}.
   ChiPFormer and WireMask-EA demonstrated a significant reduction in MacroHPWL compared to OpenROAD using TritonMacroPlacer.
   WireMask-EA achieved the best performance in terms of MacroHPWL. 
   While these AI-based placement algorithms showed good performance on several intermediate metrics, they perform poorly in terms of the end-to-end metrics compared to OpenROAD, particularly in Power, TNS, and Area.
   This outcome revealed a significant gap between the originally optimized MacroHPWL intermediate metrics and the final design PPA.

   \begin{table}[t]
   \centering
   \caption{The evaluation results of AI-based macro placement algorithms. MacroHPWL, HPWL, Congestion are intermediate metrics, and the other metrics evaluate the end-to-end performance.}
   \resizebox{0.99\textwidth}{!}{
   \begin{tabular}{@{}cccccccccc@{}}
   \toprule
   \toprule
   
   \multirow{2}{*}{\textbf{Method}} & \multicolumn{3}{c}{\textbf{Intermediate Metrics}} & \multicolumn{6}{c}{\textbf{PPA Metrics}} \\ \cmidrule(lr){2-4} \cmidrule(lr){5-10}
           & \textbf{MacroHPWL} $\downarrow$ & \textbf{HPWL} $\downarrow$ & \textbf{Congestion} $\downarrow$ &\textbf{Wirelength}  $\downarrow$ & \textbf{Power} $\downarrow$ & \textbf{WNS} $\downarrow$ & \textbf{TNS} $\downarrow$ & \textbf{NVP} $\downarrow$ & \textbf{Area} $\downarrow$ \\ \midrule
   \textbf{WireMask-EA} & \textbf{0.647}                         & 1.027                             & 1.105                                   & 1.099                                   & 1.015                              & 1.085                            & 0.995                            & 0.967                            & 1.004                             \\
   \textbf{SA}          & 0.836                                  & 1.044                             & 1.099                                   & 1.097                                   & 1.062                              &  1.121                   & 1.311                           & 1.109                            & 1.013                             \\
   \textbf{DREAMPlace}  & 0.857                                  & 0.974                             & 1.049                                   & 1.059                                   & 1.015                              & 1.112                            & 1.025                            & 1.038                            & 0.999                             \\
   \textbf{AutoDMP}     & 0.698                                  & \textbf{0.892}                    & \textbf{0.950}                          & \textbf{0.950}                          & 1.013                              & 1.196                            & 1.540                            & 1.176                            & 1.002                             \\
   \textbf{MaskPlace}   & 1.681                                  & 1.119                             & 1.148                                   & 1.148                                   & 1.051                              & 1.014                            & \textbf{0.978}                   & \textbf{0.903}                   & 1.014                             \\
   \textbf{ChiPFormer}  & 0.681                                  & 0.976                             & 1.027                                   & 1.024                                   & 1.015                              & 1.031                            & 1.355                            & 1.223                            & \textbf{0.981}                   \\ \midrule
   \textbf{OpenROAD}    & 1.000                                  & 1.000                             & 1.000                                   & 1.000                                   & \textbf{1.000}                     & \textbf{1.000}                            & 1.000                            & 1.000                            & 1.000      \\ \bottomrule  
   \end{tabular}
   }
   \label{tab_main_results_macro}
   \end{table}

   \begin{table}[t]
   \centering
   \caption{The evaluation results of AI-based standard cell placement algorithms.}
   \resizebox{0.99\textwidth}{!}{
   \begin{tabular}{@{}cccccccccc@{}}
   \toprule
   \toprule

   \multirow{2}{*}{\textbf{Method}} & \multicolumn{2}{c}{\textbf{Intermediate Metrics}} & \multicolumn{6}{c}{\textbf{PPA Metrics}} \\ \cmidrule(lr){2-3} \cmidrule(lr){4-9}
   &  \textbf{HPWL} $\downarrow$ & \textbf{Congestion} $\downarrow$ & \textbf{Wirelength} $\downarrow$ & \textbf{Power} $\downarrow$ & \textbf{WNS} $\downarrow$ & \textbf{TNS} $\downarrow$ & \textbf{NVP} $\downarrow$ & \textbf{Area} $\downarrow$ \\ \midrule
   \textbf{DREAMPlace} & \textbf{0.981}                    & \textbf{0.999}                          & 1.008                                   & \textbf{0.987}                     & 1.321                            & 4.678                            & 5.313                            & 0.996                             \\
   \textbf{AutoDMP}    & 1.124                             & 1.123                                   & 1.138                                   & 1.011                              & 1.540                            & 1.916                            & 1.119                            & \textbf{0.995}                    \\ \midrule
   \textbf{OpenROAD}   & 1.000                             & 1.000                                   & \textbf{1.000}                          & 1.000                              & \textbf{1.000}                   & \textbf{1.000}                   & \textbf{1.000 }                  & 1.000                       \\ \bottomrule   
   \end{tabular}
   }
   \label{tab:main_results_cell}
   \end{table}

   \textbf{Cell Placement} As shown in Table \ref{tab:main_results_cell}, DREAMPlace achieved the best results in the intermediate metrics of HPWL, and performed well in terms of Power. However, OpenROAD achieved the best results in WNS, TNS, and NVP, further demonstrating the inconsistency between intermediate metrics and final PPA results.

   \subsection{Correlation Analysis}
   \label{sec:analysis}
   In this section, we conduct compute and discuss the correlation between the core optimization indicator MacroHPWL, used in existing placement algorithms, and the final chip performance metrics such as WNS, TNS, and wirelength, obtained through the OpenROAD process.
   
   We use the Pearson correlation coefficient~\citep{cohen2009pearson} to evaluate the strength of linear correlation between pairs of metrics.
   The formula for calculating the Pearson correlation coefficient takes the form of
   \begin{equation}
   r = \frac{\sum (X_i - \overline{X})(Y_i - \overline{Y})}{\sqrt{\sum (X_i - \overline{X})^2 \sum (Y_i - \overline{Y})^2}},
   \end{equation}
   where \(X_i\) and \(Y_i\) are the observations, and \(\overline{X}\) and \(\overline{Y}\) are the respective means.

\begin{wrapfigure}{r}{0.4\textwidth}
       \centering
       \includegraphics[width=0.4\textwidth]{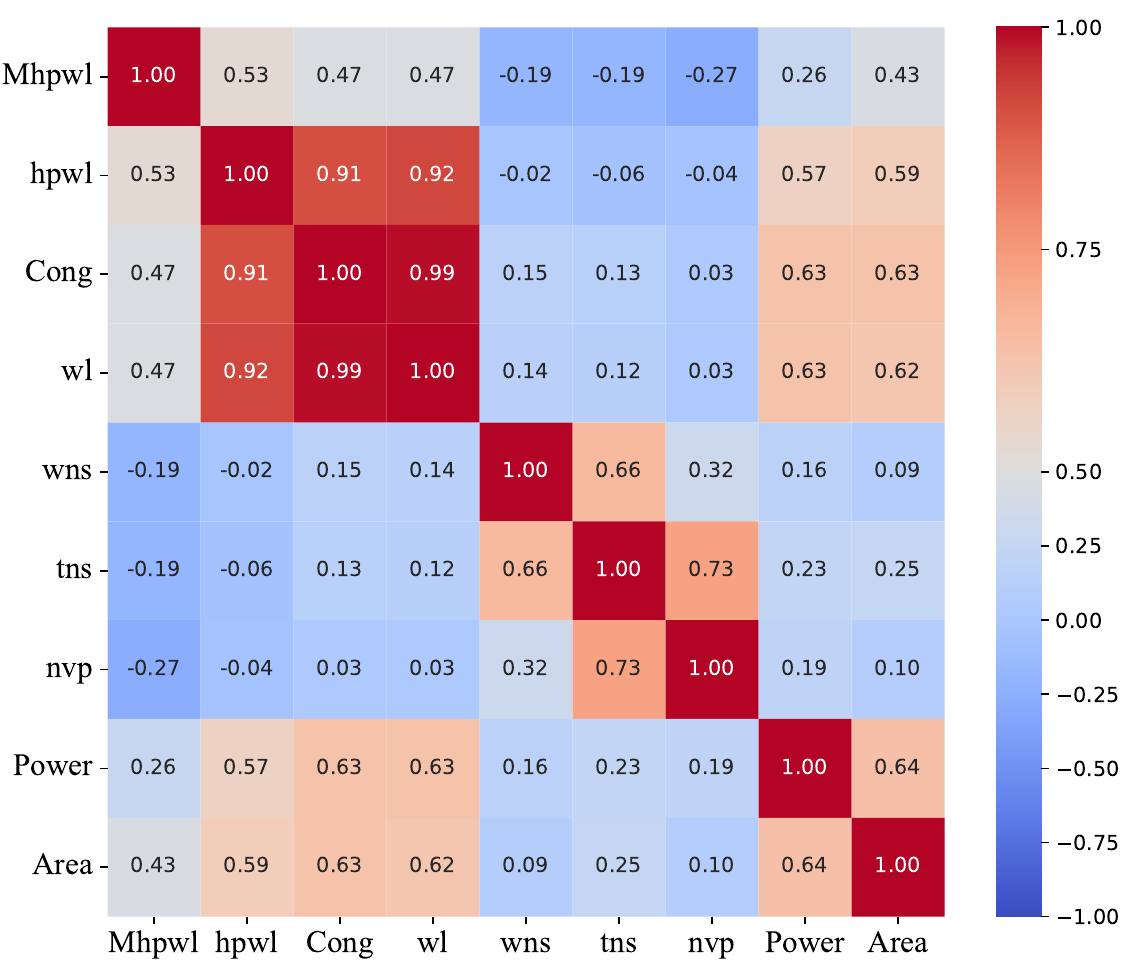}
       \vspace{-2mm}
       \caption{Correlation Between Metrics}
       \vspace{-2mm}
       \label{fig:correlation_data_final}
   \end{wrapfigure}

   The results are shown in Figure~\ref{fig:correlation_data_final}.
   To calculate the correlation, the signs of all values are adjusted so that for all metrics the lower indicates the better.
   The results show that MacroHPWL only has a weak correlation with the Wirelength, which indicates that existing algorithms that optimize MacroHPWL do not lead to an optimization on the Wirelength.
   In contrast, HPWL shows a very strong positive correlation with actual Wirelength, indicating that HPWL works as an effective surrogate for approximating the Wirelength.

   \begin{figure}
       \centering
       %\subcaptionsetup{font=scriptsize}
       \begin{subfigure}[b]{0.5\textwidth}
           \centering
           \includegraphics[width=\textwidth]{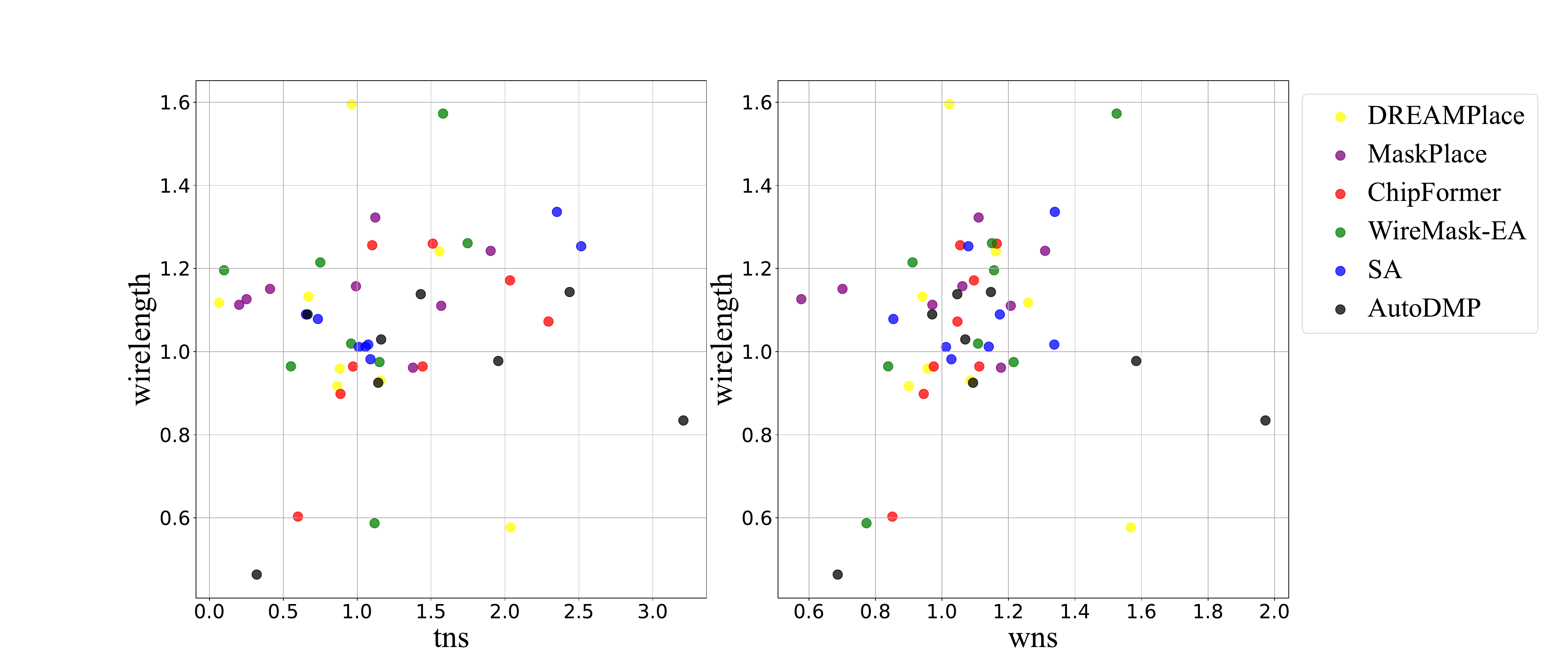}
           \caption{WL-TNS/WNS based on Method}
           \label{fig:WL-TNS/WNS based on Method}
       \end{subfigure}%
       \begin{subfigure}[b]{0.5\textwidth}
           \centering
           \includegraphics[width=\textwidth]{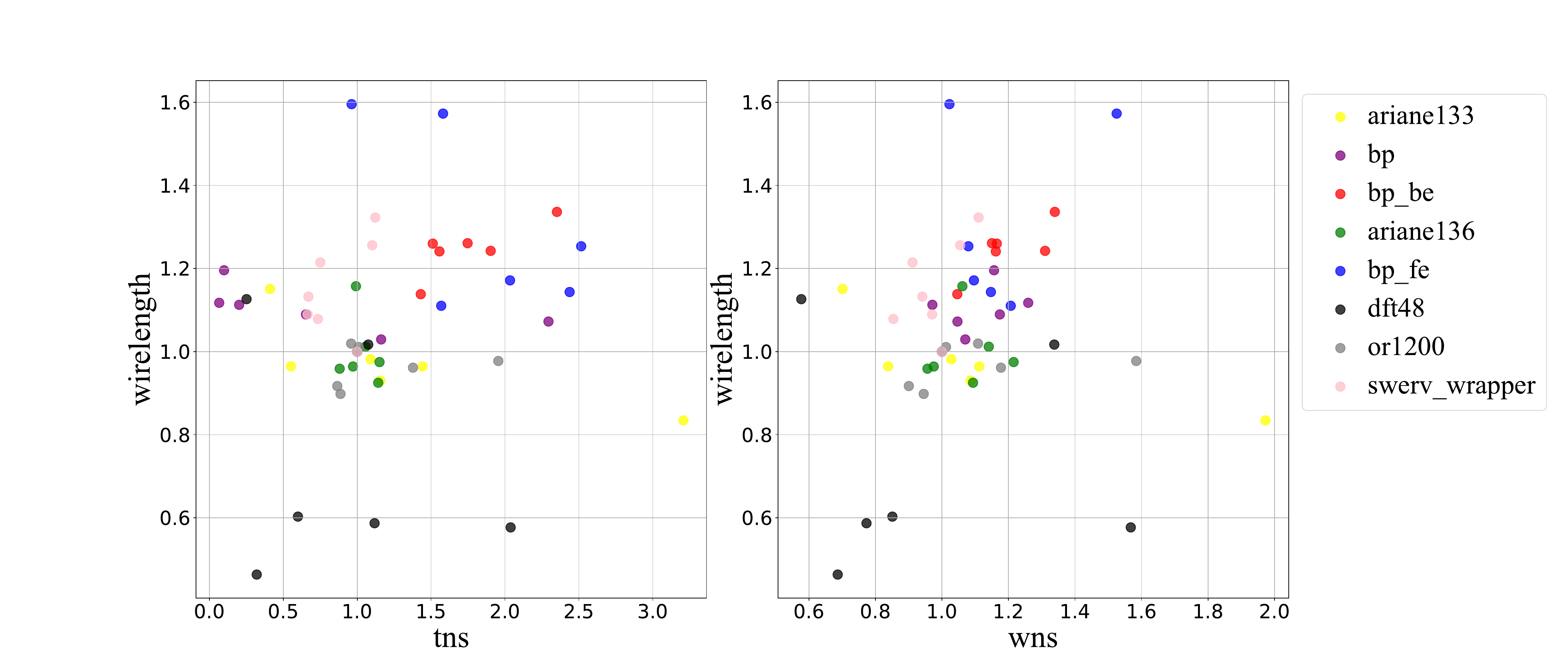}
           \caption{WL-TNS/WNS based on Case}
           \label{fig:WL-TNS/WNS based on Case}
       \end{subfigure}%

       \caption{Correlations Between Wirelength and TNS/WNS, In the visualizations, points that share the same color represent data from (a) same method or (b) same case, respectively.}
       \label{fig:correlations_wns_wirelength}
   \end{figure}
   % Additionally, the final PPA metrics of the chip also relate to WNS/TNS and congestion. The analysis shows that MacroHPWL has weak correlations with them, suggesting that optimization of MacroHPWL hardly impacts these performance metrics. Moreover wirelength also has weak correlations with all final metrics. Even if the single-point algorithm can optimize these intermediate metrics correctly, the final physical implementation result is not necessarily be optimal. Therefore, a more appropriate intermediate metric is needed, which has a higher correlation to the actual PPA objectives or can accurately reflect it.
   
   In addition to Wirelength, the final PPA metrics of the chip are associated with WNS/TNS, Area, and Power.
   The analyses shows that the correlation between MacroHPWL and these metrics is weak, indicating that optimization of MacroHPWL has minimal impact on these performance indicators.
   Moreover, the results in Figure \ref{fig:correlations_wns_wirelength} show that Wirelength exhibits weak correlations with WNS and TNS as well. 
   % Furthermore, aside from a very strong positive correlation with the final area, wirelength exhibits weak correlations with the other final metrics. 
   This implies that even if a single-point algorithm successfully optimizes metrics such as Wirelength, the ultimate physical implementation might only enhance one aspect of the PPA metrics and may not effectively optimize the other dimensions.
   Therefore, more appropriate intermediate metrics are needed to better correlate with the actual PPA objectives.
\subsection{Further Analysis}
\begin{figure}[t]
       \centering
       \includegraphics[width=0.90\textwidth]{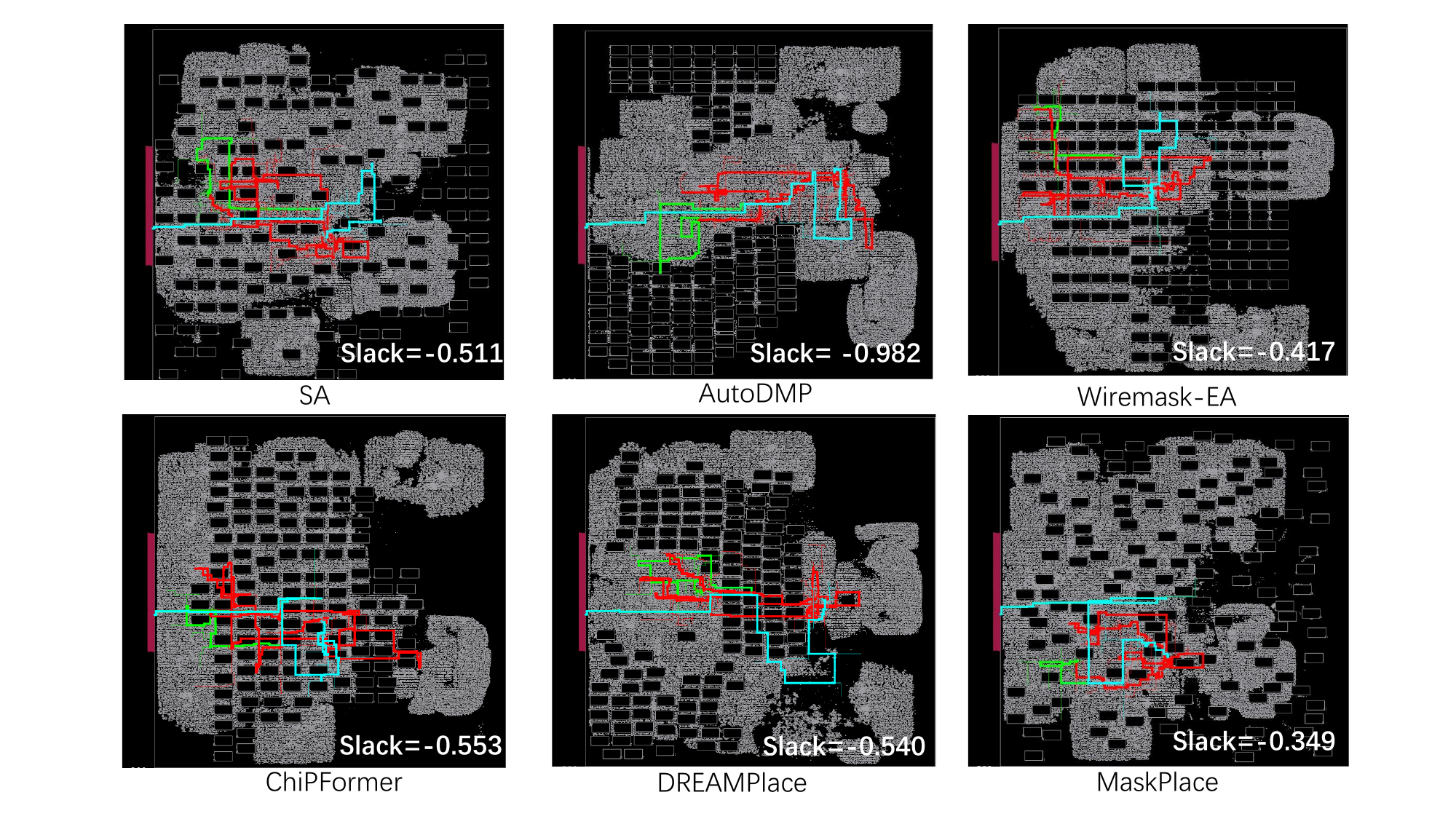}
       
       \caption{The images of the worst path for each method in ariane133.}
       \label{fig:ariane133_timing}
   \end{figure}

In this section, we analyze specific cases to assess the impact of different methods on PPA using ariane133 as the benchmark. Table~\ref{tab:ariane133_before} presents the experimental data, while Figure~\ref{fig:ariane133_timing} shows the performance variations on the worst path caused by different placement algorithms.
The AutoDMP method reduces wirelength but worsens timing, with smaller area and lower power. This is due to fewer buffers being added during timing repair, leading to degraded timing despite the reduced area and power. Moreover, wirelength and timing are not always proportional—critical paths still experience delays, indicating that global wirelength optimization may miss timing-sensitive paths, especially in worst-case scenarios. For more details, see Appendix~\ref{detailed_anal}.

% In this section, we focus on  specific cases  to analyze the impact of different methods on the final PPA. Using ariane133 as the experimental benchmark, we visualize how various methods lead to different layouts and their respective effects on the worst path. The experimental data is presented in Table~\ref{tab:ariane133_before}, and Figure~\ref{fig:ariane133_timing} highlights the performance variations on the worst path caused by different placement algorithms.
   \begin{table}[t]
       \centering
       \caption{The evaluation results of ariane133 under AI-based macro placement algorithms.}
       \resizebox{0.99\textwidth}{!}{%
       \begin{tabular}{@{{}}cccccccccc@{{}}}
   \toprule
   \multirow{2}{*}{\textbf{Method}} & \multicolumn{3}{c}{\textbf{Intermediate Metrics}} & \multicolumn{6}{c}{\textbf{PPA Metrics}} \\ \cmidrule(lr){2-4} \cmidrule(lr){5-10}

    & \textbf{MacroHPWL} $\downarrow$ & \textbf{HPWL} $\downarrow$ & \textbf{Congestion} $\downarrow$ & \textbf{Wirelength} $\downarrow$ & \textbf{Power} $\downarrow$ & \textbf{WNS} $\uparrow$ & \textbf{TNS} $\uparrow$ & \textbf{NVP} $\downarrow$ & \textbf{Area} $\downarrow$ \\
   \midrule
   \textbf{WireMask-EA} & 1124169 & 5065453.5 & 0.226 & 6583143 & 0.369 & -0.417 & -329.353 & 1970 & 349228 \\
   \textbf{SA} & 1683330 & 5187015.8 & 0.230 & 6699484 & 0.365 & -0.512 & -650.399 & 2317 & 347043 \\
   \textbf{MaskPlace} & 4444289 & 6253554.3 & 0.265 & 7853892 & 0.373 & \textbf{-0.349} & \textbf{-244.936} & \textbf{1636} & 357322 \\
   \textbf{ChiPFormer} & 1253799 & 5019138.5 & 0.226 & 6581086 & 0.370 & -0.553 & -860.952 & 2703 & 349005 \\
   \textbf{DREAMPlace} & 1111023 & 4826654.0 & 0.214 & 6348638 & 0.367 & -0.540 & -690.266 & 2307 & 348180 \\
   \textbf{AutoDMP} & \textbf{828592} & \textbf{4250870.4} & \textbf{0.192} & \textbf{5694373} & \textbf{0.350} & -0.982 & -1913.660 & 3310 & \textbf{344091} \\
   \textbf{OpenROAD} & 2685856 & 5260791.9 & 0.235 & 6825071 & 0.359 & -0.498 & -596.718 & 2274 & 352706 \\
   \bottomrule
   \end{tabular}
   
       }
       \label{tab:ariane133_before}
   \end{table}
   
   \subsection{Discussion}
   \label{sec:discussion}
   Our benchmark comprises design kits essential for each stage of the physical design flow, including netlists, libraries, rules, and constraints needed during the physical implementation stage.
   This thorough inclusion allows for a convenient and detailed evaluation of algorithms at specific stages of the physical design flow, enabling researchers and practitioners to test and compare the effectiveness of their solutions in a realistic, end-to-end environment.
   We call on the AI researchers to pay more attention on the ``shift-left'' challenge from the real-world industrial scenarios, keeping towards the mission of bridging the huge gap between academic research and industrial applications.
   
   % \subsubsection{Data Augmentation} 
   
   % Our dataset currently has limitations in terms of data volume. In the future, we aim to increase the dataset's size and include more cases from various domains. This will enhance its generalizability and robustness, making it a more comprehensive and valuable resource for researchers . By expanding the dataset, we will improve its applicability across different areas of chip design and EDA, further refining and perfecting this important resource.
   
   \section{Conclusion}
   This paper presents a comprehensive dataset that spans the entire spectrum of the EDA design process and an end-to-end evaluation method, which we used to assess several placement algorithms: SA, WireMask-EA, DREAMPlace, AutoDMP, MaskPlace, and ChiPFormer. Our evaluation revealed inconsistencies between the metrics currently emphasized by mainstream placement algorithms and the final performance outcomes. These findings highlight the need for a new perspective in the development of placement algorithms.

   % \bibliographystyle{iclr2024_conference}
   % \bibliography{iclr2024_conference}

   % \begin{thebibliography}{99}  
   % \bibitem{ref1}\href{https://github.com/Tommydag/CAN-Bus-Controller}{https://github.com/Tommydag/CAN-Bus-Controller}
   % \bibitem{ref2}\href{https://github.com/brmcfarl/iot_shield}{https://github.com/brmcfarl/iot_shield}
   % \bibitem{ref3}\href{https://github.com/WangXuan95/FPGA-CAN}{https://github.com/WangXuan95/FPGA-CAN}
    
   % \bibitem{ref4}\href{https://github.com/Hara-Laboratory/subrisc.git}{https://github.com/Hara-Laboratory/subrisc.git}  
   
   % \bibitem{ref5}\href{https://github.com/secworks/sha256.git}{https://github.com/secworks/sha256.git}
   % \bibitem{ref6}\href{https://github.com/lajanugen/8051.git }{https://github.com/lajanugen/8051.git }
   
   % \bibitem{ref7}\href{https://github.com/AngeloJacobo/FPGA_OV7670_Camera_Interface.git}{https://github.com/AngeloJacobo/FPGA_OV7670_Camera_Interface.git}
   % \bibitem{ref8}\href{https://github.com/suntodai/FPGA_image_processing}{https://github.com/suntodai/FPGA_image_processing} 
   % \bibitem{ref9}\href{https://github.com/YosysHQ/picorv32}{https://github.com/YosysHQ/picorv32}
   % \bibitem{ref10}\href{https://github.com/olofk/serv.git}{https://github.com/olofk/serv.git}
   % \bibitem{ref11}\href{https://github.com/The-OpenROAD-Project/OpenROAD-flow-scripts}
   % {https://github.com/The-OpenROAD-Project/OpenROAD-flow-scripts}
   % \end{thebibliography}

   %%%%%%%%%%%%%%%%%%%%%%%%%%%%%%%%%%%%%%%%%%%%%%%%%%%%%%%%%%%%
   \newpage

   

   \bibliography{iclr2024_conference}
   \bibliographystyle{iclr2024_conference}
   
   \clearpage
   \appendix
   
   \section{Technical details}
   \subsection{Experimental Details}
   \label{sec:exp_detail}

   In our workflow, it starts with OpenROAD. The file format for physical design in OpenROAD is OpenDB database (.odb), which contains \texttt{LEF/DEF} information. OpenROAD can be used to convert ODB to \texttt{LEF/DEF}, or \texttt{LEF/DEF} to ODB.
   Initially, the ODB after macro placement in the OpenROAD flow is converted to \texttt{LEF/DEF} files, serving as inputs for other macro placement algorithms. Since most existing macro placement algorithms only support bookshelf format input, DREAMPlace's code is used to convert \texttt{LEF/DEF} to bookshelf (.nodes, .pl, .nets) format. It is important to note that when DREAMPlace reads the DEF file, it treats the blockages in the BLOCKAGES part as virtual macros, so when using DREAMPlace's PlaceDB in Python to read macro information, the virtual macros must be excluded.
   
   For ChiPFormer, use the default model and perform 100 online iterations to fine-tune the macro layout.For MaskPlace, run 3000 epochs.Other settings use the defaults.
   
   In the method of ChipFormer and MaskPlace, the experiments were run on an NVIDIA GeForce RTX 2080 Ti, taking one day for all cases.For the other algorithms, we used 32 CPUs (Intel(R) Xeon(R) CPU E5-2667 v4 @ 3.20GHz), with a total time expenditure of two days.
   
   Next, the macro layout (.pl) files are obtained after running ChiPFormer, MaskPlace, WireMask-EA, and SA, and then DREAMPlace is used to write the macro layout into DEF files. Since DREAMPlace does not modify the blockages in the DEF file, if blockages are defined in the DEF file, additional script modifications may be required.
   
   After obtaining the macro-placed DEF files, they are converted to ODB using OpenROAD, followed by Tapcell and Welltie insertion, PDN generation, IO place, global place, detail place, and subsequent CTS and routing.

   The performance of OpenROAD is limited; in the future, other global placement tools and detailed placement tools can be used to run the entire workflow.

   \subsection{Encountered Errors}
   Due to certain parsing bugs in OpenROAD, when converting DEF files to ODB files, it is necessary to ensure that no object names in the DEF file contain "/", otherwise, there will be issues during timing tests (issues will occur when writing and reading SPEF file).
   
   Since various algorithms are academic and cannot directly set the minimum spacing between macros (which can be set in OpenROAD), the optimization of macro positions by these algorithms may result in macros being placed too closely. When integrating back into OpenROAD, the following errors might occur:
   
   OpenROAD's limited capabilities in various stages also contribute to these errors. During PDN, "Unable to repair all channels." might be encountered.
   During the global place stage, global placement might diverge.
   During the detail place stage, detailed placement might fail (because the surrounding space of certain cells makes it impossible for OpenROAD to find space for adjustments).
   % Therefore, to run the entire flow smoothly, the size of macros was enlarged when running these algorithms to prevent macros from being placed too closely (the normal size is used for calculating proxy metrics), which might not fully exploit the algorithms' performance.
   
   The code  of ChiPFormer has certain issues and needs modifications to be applied to other cases.
   DREAMPlace directly uses the mixed-size method with LEF/DEF for running macros. 
   
   For the ICCAD 2015 dataset, it includes the lib file, LEF/DEF files, netlist file, and sdc file. However, the lib file lacks buffer definitions, preventing CTS. The lef files have incomplete layer definitions, hindering routing.Additionally, OpenROAD does not support the syntax of the ICCAD 2015 sdc file.

   %AutoDMP's code does not support the mix-size format with LEF/DEF. Instead, it supports macro placement using bookshelf format, but due to the lack of a .scl file in the bookshelf, it failed.

   \section{Limitation}
   \label{sec:limitation}
   
    Our dataset currently has limitations in terms of data volume. In the future, we aim to increase the dataset's size and include more cases from various domains. This will enhance its generalizability and robustness, making it a more comprehensive and valuable resource for researchers . By expanding the dataset, we will improve its applicability across different areas of chip design and EDA, further refining and perfecting this important resource.
   
   % \section{More }
   
   \section{More Results}
   \subsection{Additional Analysis}\label{detailed_anal}
In this section, we provide a detailed analysis of the layout results for the ariane133 benchmark, focusing on the observed relationship between wirelength, timing, area, and power. The AutoDMP method shows interesting yet seemingly contradictory outcomes—reduced wirelength with worse timing, smaller area, and lower power. To understand these results, we explore the factors that contribute to these trends and propose potential directions for placement algorithm improvements.

A key factor influencing timing in large-scale circuits is buffer insertion. In general, when the wirelength between two pins is large, signal transmission delay increases. Buffer insertion helps by breaking long interconnects into smaller segments, reducing the overall transmission delay. Furthermore, buffers enhance driving capacity by reducing the load on the signal driver, which helps lower signal delay and improve timing. However, fewer buffers result in smaller area and lower power but also lead to degraded timing performance. In the case of ariane133, the AutoDMP method inserts fewer buffers during timing repair, resulting in smaller overall area and power consumption. However, the lack of sufficient buffers leads to worse timing outcomes, as seen in the experimental data. This explains why, despite achieving minimized wirelength, the timing performance does not improve accordingly.

The relationship between wirelength and timing is often assumed to be directly proportional; however, the analysis of ariane133 demonstrates that this is not always the case. Although the total wirelength (as indicated by the HPWL) is reduced, the timing performance does not improve proportionally. In fact, congestion is lower and wirelength is minimized, but critical paths still suffer from significant delays. This suggests that global wirelength optimization does not always lead to improved timing, particularly for worst-case paths. Table~\ref{tab:133_worst} provides insights into this phenomenon by highlighting the inconsistencies between wirelength and timing slack on the worst paths. Specifically, the data reveals that longer wirelength paths do not always exhibit worse timing, indicating that the optimization strategy may not have sufficiently accounted for timing-sensitive paths.

To address these issues and improve overall placement quality, several optimization strategies can be explored. One potential solution to mitigate timing degradation is to apply weighted optimization that emphasizes critical paths. By giving more weight to timing-sensitive paths during wirelength reduction, it would be possible to achieve better alignment between wirelength optimization and timing improvement. Another approach is to adopt mixed-size placement techniques that optimize both macros and standard cells simultaneously. Since macro wirelength (macro HPWL) has a weaker correlation with backend performance metrics, this approach could lead to a more balanced placement outcome, particularly for designs where macro placement significantly impacts PPA. Additionally, machine learning models that predict PPA based on placement features could provide valuable insights during placement optimization. These models would allow the algorithm to dynamically adjust its strategy based on predicted PPA outcomes, leading to more informed trade-offs between PPA.

\begin{table}[]
\centering
       \caption{Comparison of the worst path wirelength, delay, and slack for various methods on the ariane133 layout.}
       \resizebox{1\textwidth}{!}{
\begin{tabular}{@{}cccccccc@{}}
\toprule
\textbf{Method}     & WireMask-EA & SA      & MaskPlace & ChiPFormer                               & DREAMPlace & AutoDMP        & OpenROAD \\ \midrule
\textbf{wirelength} & 3892     & 4107 & 3529  &  \textbf{5761} & 4612    & 4318        & 4121  \\
\textbf{delay}      & 4.36        & 4.57    & 4.25      & 4.49                                     & 4.54       & \textbf{4.92}  & 4.44     \\
\textbf{slack}      & -0.42       & -0.54   & -0.35     & -0.55                                    & -0.54      & \textbf{-0.98} & -0.51    \\ \bottomrule
\end{tabular}}
\label{tab:133_worst}
\end{table}

   \subsection{Raw Data}\label{app_more_results}
    All the raw data from the experiment are in Tables \ref{tab:ariane133}-\ref{tab:toygpu}.

   \begin{table}[t]
       \centering
       \caption{Detailed descriptions of our collected designs.}
       \resizebox{1.0\textwidth}{!}{
       \begin{tabular}{lll}
       \hline
           Id & Design & Description \\ \hline
           1 & 8051\cite{ref6} & FPGA implementation of the 8051 Microcontroller \\ 
           2 & ariane133 \cite{ref11} & Ariane Core \\ 
           3 & ariane136 \cite{ref11} & Ariane Core \\ 
           4 & bp \cite{ref11} & Full 64-bit RISC-V Core with Cache Coherence Directory \\ 
           5 & bp\_be \cite{ref11} & Back-end of a 64-bit RISC-V Core with Cache Coherence Directory \\ 
           6 & bp\_fe \cite{ref11} & Front-end of a 64-bit RISC-V Core with Cache Coherence Directory \\ 
           7 & CAN-Bus\cite{ref1} & A CAN bus Controller \\ 
           8 & DE2\_CCD\_edge \cite{ref8} & Image processing \\ 
           9 & dft48\cite{dft}  & DFT design                   \\ \hline
           10 & FPGA-CAN\cite{ref3} & A lightweight CAN bus controller \\ 
           11& iot shield\cite{ref2} & IoT Shield for the Intel Galileo Development Board \\ 
           12 & mor1kx \cite{ref12} &an OpenRISC processor IP core\\
           13 & or1200\cite{or1200} & OpenRISC 1200 implementation \\ \hline
           14 & OV7670\_i2c \cite{ref7} & Camera interface \\ 
           15 & picorv \cite{ref9} & A Size-Optimized RISC-V CPU \\ 
           16 & serv \cite{ref10} & An award-winning bit-serial RISC-V core \\ 
           17 & sha256\cite{ref5} & Hardware implementation of the SHA-256 cryptographic hash function \\ 
           18 & subrisc\cite{ref4} & Simple Instruction-Set Computer for IoT edge devices \\ 
           19 & swerv\_wrapper \cite{ref11} & SweRV RISC-V Core 1.1 from Western Digital \\ 
           20 & toygpu \cite{ref13} &A simple GPU on a TinyFPGA BX\\
           
           \hline
       \end{tabular}}
       \label{table2}
   \end{table}

   \begin{table}[t]
       \centering
       \caption{The results of ariane133}
       \resizebox{0.99\textwidth}{!}{%
       \begin{tabular}{@{{}}cccccccccc@{{}}}
   \toprule
   \multirow{2}{*}{\textbf{Method}} & \multicolumn{3}{c}{\textbf{Intermediate Metrics}} & \multicolumn{6}{c}{\textbf{PPA Metrics}} \\ \cmidrule(lr){2-4} \cmidrule(lr){5-10}

    & \textbf{MacroHPWL} $\downarrow$ & \textbf{HPWL} $\downarrow$ & \textbf{Congestion} $\downarrow$ & \textbf{Wirelength} $\downarrow$ & \textbf{Power} $\downarrow$ & \textbf{WNS} $\uparrow$ & \textbf{TNS} $\uparrow$ & \textbf{NVP} $\downarrow$ & \textbf{Area} $\downarrow$ \\
   \midrule
   \textbf{WireMask-EA} & 1124169 & 5065453.5 & 0.226 & 6583143 & 0.369 & -0.417 & -329.353 & 1970 & 349228 \\
   \textbf{SA} & 1683330 & 5187015.8 & 0.230 & 6699484 & 0.365 & -0.512 & -650.399 & 2317 & 347043 \\
   \textbf{MaskPlace} & 4444289 & 6253554.3 & 0.265 & 7853892 & 0.373 & \textbf{-0.349} & \textbf{-244.936} & \textbf{1636} & 357322 \\
   \textbf{ChiPFormer} & 1253799 & 5019138.5 & 0.226 & 6581086 & 0.370 & -0.553 & -860.952 & 2703 & 349005 \\
   \textbf{DREAMPlace} & 1111023 & 4826654.0 & 0.214 & 6348638 & 0.367 & -0.540 & -690.266 & 2307 & 348180 \\
   \textbf{AutoDMP} & \textbf{828592} & \textbf{4250870.4} & \textbf{0.192} & \textbf{5694373} & \textbf{0.350} & -0.982 & -1913.660 & 3310 & \textbf{344091} \\
   \textbf{OpenROAD} & 2685856 & 5260791.9 & 0.235 & 6825071 & 0.359 & -0.498 & -596.718 & 2274 & 352706 \\
   \bottomrule
   \end{tabular}
   
       }
       \label{tab:ariane133}
   \end{table}

   \begin{table}[t]
       \centering
       \caption{The results of ariane136}
       \resizebox{0.99\textwidth}{!}{%
       \begin{tabular}{@{{}}cccccccccc@{{}}}
   \toprule
   \multirow{2}{*}{\textbf{Method}} & \multicolumn{3}{c}{\textbf{Intermediate Metrics}} & \multicolumn{6}{c}{\textbf{PPA Metrics}} \\ \cmidrule(lr){2-4} \cmidrule(lr){5-10}
   
    & \textbf{MacroHPWL} $\downarrow$ & \textbf{HPWL} $\downarrow$ & \textbf{Congestion} $\downarrow$ & \textbf{Wirelength} $\downarrow$ & \textbf{Power} $\downarrow$ & \textbf{WNS} $\uparrow$ & \textbf{TNS} $\uparrow$ & \textbf{NVP} $\downarrow$ & \textbf{Area} $\downarrow$ \\
   \midrule
   \textbf{WireMask-EA} & 1181404 & 5278467.4 & 0.239 & 6945252 & 0.570 & -1.726 & -4268.610 & 3628 & 358740 \\
   \textbf{SA} & 1693986 & 5608216.9 & 0.248 & 7208805 & 0.571 & -1.620 & -3917.960 & 3487 & 359532 \\
   \textbf{MaskPlace} & 4586233 & 6544926.4 & 0.285 & 8245218 & 0.583 & -1.508 & -3677.480 & 3966 & 363704 \\
   \textbf{ChiPFormer} & 1269027 & 5244085.2 & 0.237 & 6869186 & 0.566 & -1.385 & -3603.530 & 3609 & 358773 \\
   \textbf{DREAMPlace} & 1067974 & 5202240.5 & 0.231 & 6831531 & 0.571 & \textbf{-1.358} & \textbf{-3269.220} & 4350 & \textbf{356706} \\
   \textbf{AutoDMP} & \textbf{870515} & \textbf{4993936.2} & \textbf{0.228} & \textbf{6590972} & 0.562 & -1.553 & -4236.020 & 4431 & 356714 \\
   \textbf{OpenROAD} & 3067334 & 5561870.0 & 0.246 & 7124942 & \textbf{0.546} & -1.420 & -3707.930 & \textbf{3473} & 360120 \\
   \bottomrule
   \end{tabular}
   
       }
       \label{tab:ariane136}
   \end{table}

   \begin{table}[t]
       \centering
       \caption{The results of bp}
       \resizebox{0.99\textwidth}{!}{%
       \begin{tabular}{@{{}}cccccccccc@{{}}}
   \toprule
   \multirow{2}{*}{\textbf{Method}} & \multicolumn{3}{c}{\textbf{Intermediate Metrics}} & \multicolumn{6}{c}{\textbf{PPA Metrics}} \\ \cmidrule(lr){2-4} \cmidrule(lr){5-10}
   
    & \textbf{MacroHPWL} $\downarrow$ & \textbf{HPWL} $\downarrow$ & \textbf{Congestion} $\downarrow$ & \textbf{Wirelength} $\downarrow$ & \textbf{Power} $\downarrow$ & \textbf{WNS} $\uparrow$ & \textbf{TNS} $\uparrow$ & \textbf{NVP} $\downarrow$ & \textbf{Area} $\downarrow$ \\
   \midrule
   \textbf{WireMask-EA} & \textbf{112560} & 7526021.6 & 0.437 & 10002159 & 0.254 & -1.936 & -21.812 & 326 & 490622 \\
   \textbf{SA} & 151696 & 6927899.3 & 0.398 & 9113560 & 0.253 & -1.965 & -143.317 & 969 & 489850 \\
   \textbf{MaskPlace} & 324368 & 7135790.7 & 0.407 & 9307133 & 0.253 & \textbf{-1.625} & -44.217 & 478 & 489867 \\
   \textbf{ChiPFormer} & 138943 & 6857073.6 & 0.392 & 8970666 & 0.250 & -1.752 & -502.977 & 2179 & 490305 \\
   \textbf{DREAMPlace} & 161581 & 6989990.7 & 0.409 & 9347541 & 0.250 & -2.108 & \textbf{-14.609} & \textbf{192} & 489693 \\
   \textbf{AutoDMP} & 135090 & 6568910.6 & 0.376 & 8610145 & \textbf{0.247} & -1.791 & -254.786 & 1821 & 486445 \\
   \textbf{OpenROAD} & 119123 & \textbf{6389902.6} & \textbf{0.365} & \textbf{8365647} & 0.247 & -1.674 & -219.229 & 1095 & \textbf{484494} \\
   \bottomrule
   \end{tabular}
   
       }
       \label{tab:bp}
   \end{table}

   \begin{table}[t]
       \centering
       \caption{The results of bp\_be}
       \resizebox{0.99\textwidth}{!}{%
       \begin{tabular}{@{{}}cccccccccc@{{}}}
   \toprule
   \multirow{2}{*}{\textbf{Method}} & \multicolumn{3}{c}{\textbf{Intermediate Metrics}} & \multicolumn{6}{c}{\textbf{PPA Metrics}} \\ \cmidrule(lr){2-4} \cmidrule(lr){5-10}
   
    & \textbf{MacroHPWL} $\downarrow$ & \textbf{HPWL} $\downarrow$ & \textbf{Congestion} $\downarrow$ & \textbf{Wirelength} $\downarrow$ & \textbf{Power} $\downarrow$ & \textbf{WNS} $\uparrow$ & \textbf{TNS} $\uparrow$ & \textbf{NVP} $\downarrow$ & \textbf{Area} $\downarrow$ \\
   \midrule
   \textbf{WireMask-EA} & 66729 & 2678100.1 & 0.524 & 3574875 & 0.467 & -2.142 & -4093.970 & 5131 & 123966 \\
   \textbf{SA} & 85984 & 2568796.6 & 0.557 & 3788559 & 0.459 & -2.494 & -5510.140 & 5381 & 124667 \\
   \textbf{MaskPlace} & 114314 & 2548230.4 & 0.513 & 3522611 & 0.456 & -2.439 & -4459.870 & 5173 & 124938 \\
   \textbf{ChiPFormer} & \textbf{62168} & 2525703.6 & 0.524 & 3572070 & 0.425 & -2.169 & -3541.820 & 5100 & \textbf{110904} \\
   \textbf{DREAMPlace} & 87304 & 2533359.4 & 0.516 & 3518916 & 0.458 & -2.163 & -3648.020 & 6026 & 125221 \\
   \textbf{AutoDMP} & 86263 & 2539005.8 & 0.468 & 3227167 & 0.453 & -1.948 & -3351.120 & 5045 & 125471 \\
   \textbf{OpenROAD} & 76561 & \textbf{2370091.6} & \textbf{0.409} & \textbf{2835542} & \textbf{0.409} & \textbf{-1.862} & \textbf{-2343.290} & \textbf{5013} & 118786 \\
   \bottomrule
   \end{tabular}
   
       }
       \label{tab:bp_be}
   \end{table}

   \begin{table}[t]
       \centering
       \caption{The results of bp\_fe}
       \resizebox{0.99\textwidth}{!}{%
       \begin{tabular}{@{{}}cccccccccc@{{}}}
   \toprule
   \multirow{2}{*}{\textbf{Method}} & \multicolumn{3}{c}{\textbf{Intermediate Metrics}} & \multicolumn{6}{c}{\textbf{PPA Metrics}} \\ \cmidrule(lr){2-4} \cmidrule(lr){5-10}
   
    & \textbf{MacroHPWL} $\downarrow$ & \textbf{HPWL} $\downarrow$ & \textbf{Congestion} $\downarrow$ & \textbf{Wirelength} $\downarrow$ & \textbf{Power} $\downarrow$ & \textbf{WNS} $\uparrow$ & \textbf{TNS} $\uparrow$ & \textbf{NVP} $\downarrow$ & \textbf{Area} $\downarrow$ \\
   \midrule
   \textbf{WireMask-EA} & 45985 & 1863651.1 & 0.510 & 2783740 & 0.309 & -1.666 & -777.420 & 2628 & 77648.300 \\
   \textbf{SA} & 59872 & 1603778.9 & 0.404 & 2218668 & 0.310 & -1.179 & -1236.880 & 2990 & 81919.500 \\
   \textbf{MaskPlace} & 64024 & 1540944.3 & 0.355 & 1965209 & 0.309 & -1.318 & -771.363 & 2582 & 77881.100 \\
   \textbf{ChiPFormer} & \textbf{45754} & 1544281.0 & 0.375 & 2073376 & 0.299 & -1.197 & -1000.190 & 2714 & \textbf{72723.900} \\
   \textbf{DREAMPlace} & 60279 & 1975681.6 & 0.508 & 2823861 & 0.294 & -1.117 & \textbf{-473.261} & \textbf{1849} & 77427.800 \\
   \textbf{AutoDMP} & 56107 & 1560612.6 & 0.365 & 2023847 & 0.312 & -1.253 & -1198.430 & 2626 & 81294.400 \\
   \textbf{OpenROAD} & 53842 & \textbf{1468799.8} & \textbf{0.319} & \textbf{1770176} & \textbf{0.285} & \textbf{-1.092} & -491.698 & 1870 & 75884.500 \\
   \bottomrule
   \end{tabular}
   
       }
       \label{tab:bp_fe}
   \end{table}

   \begin{table}[t]
       \centering
       \caption{The results of dft48}
       \resizebox{0.99\textwidth}{!}{%
       \begin{tabular}{@{{}}cccccccccc@{{}}}
   \toprule
   \multirow{2}{*}{\textbf{Method}} & \multicolumn{3}{c}{\textbf{Intermediate Metrics}} & \multicolumn{6}{c}{\textbf{PPA Metrics}} \\ \cmidrule(lr){2-4} \cmidrule(lr){5-10}
   
    & \textbf{MacroHPWL} $\downarrow$ & \textbf{HPWL} $\downarrow$ & \textbf{Congestion} $\downarrow$ & \textbf{Wirelength} $\downarrow$ & \textbf{Power} $\downarrow$ & \textbf{WNS} $\uparrow$ & \textbf{TNS} $\uparrow$ & \textbf{NVP} $\downarrow$ & \textbf{Area} $\downarrow$ \\
   \midrule
   \textbf{WireMask-EA} & 335246296 & 1039236.4 & 0.114 & 1330932 & 0.239 & -0.510 & -439.378 & 2035 & 81353.700 \\
   \textbf{SA} & 339003863 & 1815564.2 & 0.198 & 2305190 & 0.304 & -0.882 & -422.568 & 1546 & 84170.100 \\
   \textbf{MaskPlace} & 1409226442 & 2117408.1 & 0.223 & 2552954 & 0.258 & \textbf{-0.381} & \textbf{-98.916} & \textbf{548} & 83185.400 \\
   \textbf{ChiPFormer} & 431188529 & 1094111.0 & 0.119 & 1366989 & 0.242 & -0.561 & -235.601 & 974 & 80770.400 \\
   \textbf{DREAMPlace} & \textbf{267700115} & 800165.5 & 0.112 & 1308094 & 0.234 & -1.034 & -801.257 & 2274 & \textbf{79407.400} \\
   \textbf{AutoDMP} & 270017585 & \textbf{774418.4} & \textbf{0.091} & \textbf{1051008} & \textbf{0.232} & -0.453 & -125.948 & 797 & 80201.100 \\
   \textbf{OpenROAD} & 752297935 & 1981412.2 & 0.198 & 2267002 & 0.257 & -0.660 & -393.111 & 1411 & 84300.200 \\
   \bottomrule
   \end{tabular}
   
       }
       \label{tab:dft48}
   \end{table}

   \begin{table}[t]
       \centering
       \caption{The results of or1200}
       \resizebox{0.99\textwidth}{!}{%
       \begin{tabular}{@{{}}cccccccccc@{{}}}
   \toprule
   \multirow{2}{*}{\textbf{Method}} & \multicolumn{3}{c}{\textbf{Intermediate Metrics}} & \multicolumn{6}{c}{\textbf{PPA Metrics}} \\ \cmidrule(lr){2-4} \cmidrule(lr){5-10}
   
    & \textbf{MacroHPWL} $\downarrow$ & \textbf{HPWL} $\downarrow$ & \textbf{Congestion} $\downarrow$ & \textbf{Wirelength} $\downarrow$ & \textbf{Power} $\downarrow$ & \textbf{WNS} $\uparrow$ & \textbf{TNS} $\uparrow$ & \textbf{NVP} $\downarrow$ & \textbf{Area} $\downarrow$ \\
   \midrule
   \textbf{WireMask-EA} & 102261063 & 1196931.6 & 0.035 & 1432301 & \textbf{0.049} & -1.417 & -1530.180 & \textbf{1754} & 66547.600 \\
   \textbf{SA} & 113496323 & 1200696.8 & 0.034 & 1420640 & 0.059 & -1.294 & -1613.410 & 2674 & 66937.300 \\
   \textbf{MaskPlace} & 208736767 & 1166784.1 & 0.033 & 1350820 & 0.059 & -1.505 & -2198.370 & 2684 & 66953.500 \\
   \textbf{ChiPFormer} & \textbf{95035903} & 1089714.1 & \textbf{0.030} & \textbf{1262132} & 0.057 & -1.208 & -1415.960 & 2669 & 66256.900 \\
   \textbf{DREAMPlace} & 153596940 & 1099937.2 & 0.031 & 1288746 & 0.057 & \textbf{-1.151} & \textbf{-1380.980} & 2670 & 65981.000 \\
   \textbf{AutoDMP} & 122473700 & \textbf{1058807.4} & 0.033 & 1373410 & 0.058 & -2.025 & -3119.280 & 2760 & \textbf{65756.800} \\
   \textbf{OpenROAD} & 187077195 & 1223792.9 & 0.034 & 1405193 & 0.060 & -1.278 & -1595.730 & 2677 & 67276.500 \\
   \bottomrule
   \end{tabular}
   
       }
       \label{tab:or1200}
   \end{table}

   \begin{table}[t]
       \centering
       \caption{The results of swerv\_wrapper}
       \resizebox{0.99\textwidth}{!}{%
       \begin{tabular}{@{{}}cccccccccc@{{}}}
   \toprule
   \multirow{2}{*}{\textbf{Method}} & \multicolumn{3}{c}{\textbf{Intermediate Metrics}} & \multicolumn{6}{c}{\textbf{PPA Metrics}} \\ \cmidrule(lr){2-4} \cmidrule(lr){5-10}
   
    & \textbf{MacroHPWL} $\downarrow$ & \textbf{HPWL} $\downarrow$ & \textbf{Congestion} $\downarrow$ & \textbf{Wirelength} $\downarrow$ & \textbf{Power} $\downarrow$ & \textbf{WNS} $\uparrow$ & \textbf{TNS} $\uparrow$ & \textbf{NVP} $\downarrow$ & \textbf{Area} $\downarrow$ \\
   \midrule
   \textbf{WireMask-EA} & 474447 & 3893655.8 & 0.414 & 4854661 & 0.666 & -1.027 & -873.506 & 1518 & 205627 \\
   \textbf{SA} & 630919 & 3811960.2 & 0.368 & 4310596 & 0.648 & \textbf{-0.962} & -854.737 & 1811 & \textbf{200280} \\
   \textbf{MaskPlace} & 1270726 & 4193788.0 & 0.452 & 5286392 & 0.690 & -1.251 & -1306.460 & 1696 & 206481 \\
   \textbf{ChiPFormer} & \textbf{437638} & 4060188.1 & 0.428 & 5019849 & 0.674 & -1.189 & -1282.240 & 2139 & 205775 \\
   \textbf{DREAMPlace} & 865544 & 3575327.7 & 0.366 & 4525348 & 0.646 & -1.061 & -780.196 & 1608 & 203896 \\
   \textbf{AutoDMP} & 450662 & 3213155.0 & 0.372 & 4354901 & 0.648 & -1.094 & \textbf{-773.339} & 1529 & 200823 \\
   \textbf{OpenROAD} & 666704 & \textbf{3177043.0} & \textbf{0.339} & \textbf{3997163} & \textbf{0.628} & -1.127 & -1163.820 & \textbf{1515} & 202853 \\
   \bottomrule
   \end{tabular}
   
       }
       \label{tab:swerv_wrapper}
   \end{table}

   \begin{table}[t]
       \centering
       \caption{The results of 8051-master}
       \resizebox{0.99\textwidth}{!}{%
       \begin{tabular}{@{{}}ccccccccc@{{}}}
   \toprule
   \multirow{2}{*}{\textbf{Method}} & \multicolumn{2}{c}{\textbf{Intermediate Metrics}} & \multicolumn{6}{c}{\textbf{PPA Metrics}} \\ \cmidrule(lr){2-3} \cmidrule(lr){4-9}
    & \textbf{HPWL} $\downarrow$ & \textbf{Congestion} $\downarrow$ & \textbf{Wirelength} $\downarrow$ & \textbf{Power} $\downarrow$ & \textbf{WNS} $\uparrow$ & \textbf{TNS} $\uparrow$ & \textbf{NVP} $\downarrow$ & \textbf{Area} $\downarrow$ \\
   \midrule
   \textbf{DREAMPlace} & 142162.6 & 0.201 & 210127 & 0.075 & -0.634 & -16.460 & \textbf{35} & 29484.200 \\
   \textbf{AutoDMP} & \textbf{131174.7} & \textbf{0.186} & \textbf{195536} & 0.074 & -0.633 & \textbf{-15.643} & \textbf{35} & \textbf{29365.600} \\
   \textbf{OpenROAD} & 141245.3 & 0.201 & 207737 & \textbf{0.073} & \textbf{-0.603} & -16.664 & 41 & 29461.400 \\
   \bottomrule
   \end{tabular}
   
       }
       \label{tab:8051-master}
   \end{table}

   \begin{table}[t]
       \centering
       \caption{The results of CAN-Bus}
       \resizebox{0.99\textwidth}{!}{%
       \begin{tabular}{@{{}}ccccccccc@{{}}}
   \toprule
   \multirow{2}{*}{\textbf{Method}} & \multicolumn{2}{c}{\textbf{Intermediate Metrics}} & \multicolumn{6}{c}{\textbf{PPA Metrics}} \\ \cmidrule(lr){2-3} \cmidrule(lr){4-9}
   
    & \textbf{HPWL} $\downarrow$ & \textbf{Congestion} $\downarrow$ & \textbf{Wirelength} $\downarrow$ & \textbf{Power} $\downarrow$ & \textbf{WNS} $\uparrow$ & \textbf{TNS} $\uparrow$ & \textbf{NVP} $\downarrow$ & \textbf{Area} $\downarrow$ \\
   \midrule
   \textbf{DREAMPlace} & \textbf{3916.8} & 0.093 & 5683 & \textbf{0.005} & -0.074 & -0.074 & \textbf{1} & 1675.270 \\
   \textbf{AutoDMP} & 4210.8 & 0.097 & 6024 & 0.005 & \textbf{-0.074} & \textbf{-0.074} & \textbf{1} & 1668.880 \\
   \textbf{OpenROAD} & 3955.1 & \textbf{0.092} & \textbf{5623} & 0.005 & -0.079 & -0.079 & \textbf{1} & \textbf{1665.430} \\
   \bottomrule
   \end{tabular}
   
       }
       \label{tab:CAN-Bus}
   \end{table}

   \begin{table}[t]
       \centering
       \caption{The results of DE2\_CCD\_edge}
       \resizebox{0.99\textwidth}{!}{%
       \begin{tabular}{@{{}}ccccccccc@{{}}}
   \toprule
   \multirow{2}{*}{\textbf{Method}} & \multicolumn{2}{c}{\textbf{Intermediate Metrics}} & \multicolumn{6}{c}{\textbf{PPA Metrics}} \\ \cmidrule(lr){2-3} \cmidrule(lr){4-9}
   
    & \textbf{HPWL} $\downarrow$ & \textbf{Congestion} $\downarrow$ & \textbf{Wirelength} $\downarrow$ & \textbf{Power} $\downarrow$ & \textbf{WNS} $\uparrow$ & \textbf{TNS} $\uparrow$ & \textbf{NVP} $\downarrow$ & \textbf{Area} $\downarrow$ \\
   \midrule
   \textbf{DREAMPlace} & \textbf{16556.5} & 0.123 & 23453 & 0.201 & -0.924 & -184.592 & 736 & 5797.470 \\
   \textbf{AutoDMP} & 22203.0 & 0.160 & 30745 & 0.211 & -0.930 & -200.125 & 710 & 5812.630 \\
   \textbf{OpenROAD} & 16689.0 & \textbf{0.120} & \textbf{22746} & \textbf{0.199} & \textbf{-0.919} & \textbf{-165.009} & \textbf{609} & \textbf{5758.100} \\
   \bottomrule
   \end{tabular}
   
       }
       \label{tab:DE2_CCD_edge}
   \end{table}

   \begin{table}[t]
       \centering
       \caption{The results of FPGA-CAN}
       \resizebox{0.99\textwidth}{!}{%
       \begin{tabular}{@{{}}ccccccccc@{{}}}
   \toprule
   \multirow{2}{*}{\textbf{Method}} & \multicolumn{2}{c}{\textbf{Intermediate Metrics}} & \multicolumn{6}{c}{\textbf{PPA Metrics}} \\ \cmidrule(lr){2-3} \cmidrule(lr){4-9}
   
    & \textbf{HPWL} $\downarrow$ & \textbf{Congestion} $\downarrow$ & \textbf{Wirelength} $\downarrow$ & \textbf{Power} $\downarrow$ & \textbf{WNS} $\uparrow$ & \textbf{TNS} $\uparrow$ & \textbf{NVP} $\downarrow$ & \textbf{Area} $\downarrow$ \\
   \midrule
   \textbf{DREAMPlace} & 1472841.9 & 0.181 & 2437446 & \textbf{0.738} & -1.027 & -678.865 & 17688 & \textbf{378643} \\
   \textbf{AutoDMP} & \textbf{1437916.7} & \textbf{0.178} & \textbf{2393640} & 0.745 & -0.892 & -108.201 & 3864 & 379675 \\
   \textbf{OpenROAD} & 1483498.8 & 0.180 & 2394544 & 0.742 & \textbf{-0.404} & \textbf{-30.331} & \textbf{1843} & 379193 \\
   \bottomrule
   \end{tabular}
   
       }
       \label{tab:FPGA-CAN}
   \end{table}

   \begin{table}[t]
       \centering
       \caption{The results of OV7670\_i2c}
       \resizebox{0.99\textwidth}{!}{%
       \begin{tabular}{@{{}}ccccccccc@{{}}}
   \toprule
   \multirow{2}{*}{\textbf{Method}} & \multicolumn{2}{c}{\textbf{Intermediate Metrics}} & \multicolumn{6}{c}{\textbf{PPA Metrics}} \\ \cmidrule(lr){2-3} \cmidrule(lr){4-9}
   
    & \textbf{HPWL} $\downarrow$ & \textbf{Congestion} $\downarrow$ & \textbf{Wirelength} $\downarrow$ & \textbf{Power} $\downarrow$ & \textbf{WNS} $\uparrow$ & \textbf{TNS} $\uparrow$ & \textbf{NVP} $\downarrow$ & \textbf{Area} $\downarrow$ \\
   \midrule
   \textbf{DREAMPlace} & 733113.0 & 0.215 & 1164763 & 0.625 & -0.579 & -82.237 & 3154 & 157496 \\
   \textbf{AutoDMP} & \textbf{701777.0} & \textbf{0.201} & \textbf{1091553} & \textbf{0.612} & -0.561 & \textbf{-23.030} & \textbf{1053} & \textbf{155624} \\
   \textbf{OpenROAD} & 723012.1 & 0.208 & 1127120 & 0.617 & \textbf{-0.466} & -30.612 & 2327 & 156210 \\
   \bottomrule
   \end{tabular}
   
       }
       \label{tab:OV7670_i2c}
   \end{table}

   \begin{table}[t]
       \centering
       \caption{The results of iot\_shield}
       \resizebox{0.99\textwidth}{!}{%
       \begin{tabular}{@{{}}ccccccccc@{{}}}
   \toprule
   \multirow{2}{*}{\textbf{Method}} & \multicolumn{2}{c}{\textbf{Intermediate Metrics}} & \multicolumn{6}{c}{\textbf{PPA Metrics}} \\ \cmidrule(lr){2-3} \cmidrule(lr){4-9}
   
    & \textbf{HPWL} $\downarrow$ & \textbf{Congestion} $\downarrow$ & \textbf{Wirelength} $\downarrow$ & \textbf{Power} $\downarrow$ & \textbf{WNS} $\uparrow$ & \textbf{TNS} $\uparrow$ & \textbf{NVP} $\downarrow$ & \textbf{Area} $\downarrow$ \\
   \midrule
   \textbf{DREAMPlace} & \textbf{5014.3} & \textbf{0.118} & \textbf{7346} & \textbf{0.006} & -0.167 & -2.457 & 27 & \textbf{1712.770} \\
   \textbf{AutoDMP} & 5976.2 & 0.137 & 8653 & 0.006 & \textbf{-0.152} & -2.367 & 29 & 1730.600 \\
   \textbf{OpenROAD} & 5067.3 & 0.119 & 7349 & 0.006 & -0.152 & \textbf{-2.186} & \textbf{20} & 1719.420 \\
   \bottomrule
   \end{tabular}
   
       }
       \label{tab:iot_shield}
   \end{table}

   \begin{table}[t]
       \centering
       \caption{The results of mor1kx}
       \resizebox{0.99\textwidth}{!}{%
       \begin{tabular}{@{{}}ccccccccc@{{}}}
   \toprule
   \multirow{2}{*}{\textbf{Method}} & \multicolumn{2}{c}{\textbf{Intermediate Metrics}} & \multicolumn{6}{c}{\textbf{PPA Metrics}} \\ \cmidrule(lr){2-3} \cmidrule(lr){4-9}
   
    & \textbf{HPWL} $\downarrow$ & \textbf{Congestion} $\downarrow$ & \textbf{Wirelength} $\downarrow$ & \textbf{Power} $\downarrow$ & \textbf{WNS} $\uparrow$ & \textbf{TNS} $\uparrow$ & \textbf{NVP} $\downarrow$ & \textbf{Area} $\downarrow$ \\
   \midrule
   \textbf{DREAMPlace} & \textbf{1723145.8} & \textbf{0.366} & \textbf{3421791} & \textbf{0.703} & -1.863 & -19564.300 & 26568 & 289570 \\
   \textbf{AutoDMP} & 2258941.0 & 0.524 & 4893199 & 0.735 & -3.879 & -44135.900 & 26600 & \textbf{285455} \\
   \textbf{OpenROAD} & 2006747.0 & 0.393 & 3661548 & 0.703 & \textbf{-0.796} & \textbf{-4928.920} & \textbf{25509} & 313065 \\
   \bottomrule
   \end{tabular}
   
       }
       \label{tab:mor1kx}
   \end{table}

   \begin{table}[t]
       \centering
       \caption{The results of picorv32}
       \resizebox{0.99\textwidth}{!}{%
       \begin{tabular}{@{{}}ccccccccc@{{}}}
   \toprule
   \multirow{2}{*}{\textbf{Method}} & \multicolumn{2}{c}{\textbf{Intermediate Metrics}} & \multicolumn{6}{c}{\textbf{PPA Metrics}} \\ \cmidrule(lr){2-3} \cmidrule(lr){4-9}
   
    & \textbf{HPWL} $\downarrow$ & \textbf{Congestion} $\downarrow$ & \textbf{Wirelength} $\downarrow$ & \textbf{Power} $\downarrow$ & \textbf{WNS} $\uparrow$ & \textbf{TNS} $\uparrow$ & \textbf{NVP} $\downarrow$ & \textbf{Area} $\downarrow$ \\
   \midrule
   \textbf{DREAMPlace} & 88407.3 & \textbf{0.194} & 135544 & \textbf{0.048} & -0.147 & \textbf{-0.521} & \textbf{29} & \textbf{18951.400} \\
   \textbf{AutoDMP} & 132458.3 & 0.244 & 170435 & 0.050 & -0.142 & -0.525 & 37 & 19039.000 \\
   \textbf{OpenROAD} & \textbf{87168.8} & 0.195 & \textbf{134211} & 0.048 & \textbf{-0.107} & -0.661 & 38 & 18972.700 \\
   \bottomrule
   \end{tabular}
   
       }
       \label{tab:picorv32}
   \end{table}

   \begin{table}[t]
       \centering
       \caption{The results of serv}
       \resizebox{0.99\textwidth}{!}{%
       \begin{tabular}{@{{}}ccccccccc@{{}}}
   \toprule
   \multirow{2}{*}{\textbf{Method}} & \multicolumn{2}{c}{\textbf{Intermediate Metrics}} & \multicolumn{6}{c}{\textbf{PPA Metrics}} \\ \cmidrule(lr){2-3} \cmidrule(lr){4-9}
   
    & \textbf{HPWL} $\downarrow$ & \textbf{Congestion} $\downarrow$ & \textbf{Wirelength} $\downarrow$ & \textbf{Power} $\downarrow$ & \textbf{WNS} $\uparrow$ & \textbf{TNS} $\uparrow$ & \textbf{NVP} $\downarrow$ & \textbf{Area} $\downarrow$ \\
   \midrule
   \textbf{DREAMPlace} & 7194.0 & 0.165 & 11128 & 0.007 & \textbf{-0.563} & -10.333 & 81 & 2125.870 \\
   \textbf{AutoDMP} & 10361.9 & 0.261 & 17455 & 0.008 & -0.593 & -11.893 & 120 & 2193.170 \\
   \textbf{OpenROAD} & \textbf{7165.0} & \textbf{0.163} & \textbf{10638} & \textbf{0.007} & -0.565 & \textbf{-10.210} & \textbf{73} & \textbf{2097.940} \\
   \bottomrule
   \end{tabular}
   
       }
       \label{tab:serv}
   \end{table}

   \begin{table}[t]
       \centering
       \caption{The results of sha256}
       \resizebox{0.99\textwidth}{!}{%
       \begin{tabular}{@{{}}ccccccccc@{{}}}
   \toprule
   \multirow{2}{*}{\textbf{Method}} & \multicolumn{2}{c}{\textbf{Intermediate Metrics}} & \multicolumn{6}{c}{\textbf{PPA Metrics}} \\ \cmidrule(lr){2-3} \cmidrule(lr){4-9}
   
    & \textbf{HPWL} $\downarrow$ & \textbf{Congestion} $\downarrow$ & \textbf{Wirelength} $\downarrow$ & \textbf{Power} $\downarrow$ & \textbf{WNS} $\uparrow$ & \textbf{TNS} $\uparrow$ & \textbf{NVP} $\downarrow$ & \textbf{Area} $\downarrow$ \\
   \midrule
   \textbf{DREAMPlace} & 114376.4 & 0.203 & 170230 & \textbf{0.161} & -0.542 & -9.367 & 40 & 23476.400 \\
   \textbf{AutoDMP} & 120832.7 & 0.206 & 173822 & 0.172 & -0.573 & -9.891 & \textbf{39} & \textbf{23398.700} \\
   \textbf{OpenROAD} & \textbf{113841.8} & \textbf{0.202} & \textbf{168199} & 0.162 & \textbf{-0.528} & \textbf{-8.667} & \textbf{39} & 23426.900 \\
   \bottomrule
   \end{tabular}
   
       }
       \label{tab:sha256}
   \end{table}

   \begin{table}[t]
       \centering
       \caption{The results of subrisc}
       \resizebox{0.99\textwidth}{!}{%
       \begin{tabular}{@{{}}ccccccccc@{{}}}
   \toprule
   \multirow{2}{*}{\textbf{Method}} & \multicolumn{2}{c}{\textbf{Intermediate Metrics}} & \multicolumn{6}{c}{\textbf{PPA Metrics}} \\ \cmidrule(lr){2-3} \cmidrule(lr){4-9}
   
    & \textbf{HPWL} $\downarrow$ & \textbf{Congestion} $\downarrow$ & \textbf{Wirelength} $\downarrow$ & \textbf{Power} $\downarrow$ & \textbf{WNS} $\uparrow$ & \textbf{TNS} $\uparrow$ & \textbf{NVP} $\downarrow$ & \textbf{Area} $\downarrow$ \\
   \midrule
   \textbf{DREAMPlace} & 13920769.0 & 0.264 & 21051063 & 2.774 & -0.847 & -5742.680 & 29283 & 2341490 \\
   \textbf{AutoDMP} & \textbf{13446106.8} & \textbf{0.238} & \textbf{19114559} & \textbf{2.697} & \textbf{-0.673} & \textbf{-150.178} & \textbf{466} & \textbf{2319150} \\
   \textbf{OpenROAD} & 14386152.9 & 0.253 & 20108941 & 3.425 & -0.782 & -305.819 & 674 & 2334330 \\
   \bottomrule
   \end{tabular}
   
       }
       \label{tab:subrisc}
   \end{table}

   \begin{table}[t]
       \centering
       \caption{The results of toygpu}
       \resizebox{0.99\textwidth}{!}{%
       \begin{tabular}{@{{}}ccccccccc@{{}}}
   \toprule
   \multirow{2}{*}{\textbf{Method}} & \multicolumn{2}{c}{\textbf{Intermediate Metrics}} & \multicolumn{6}{c}{\textbf{PPA Metrics}} \\ \cmidrule(lr){2-3} \cmidrule(lr){4-9}
   
    & \textbf{HPWL} $\downarrow$ & \textbf{Congestion} $\downarrow$ & \textbf{Wirelength} $\downarrow$ & \textbf{Power} $\downarrow$ & \textbf{WNS} $\uparrow$ & \textbf{TNS} $\uparrow$ & \textbf{NVP} $\downarrow$ & \textbf{Area} $\downarrow$ \\
   \midrule
   \textbf{DREAMPlace} & \textbf{4347733.8} & 0.198 & 6784639 & \textbf{1.101} & -1.503 & -131.661 & \textbf{101} & 979293 \\
   \textbf{AutoDMP} & 4444860.0 & \textbf{0.194} & \textbf{6595166} & 1.109 & -2.418 & -201.797 & 107 & \textbf{967780} \\
   \textbf{OpenROAD} & 4657703.6 & 0.212 & 7185236 & 1.108 & \textbf{-1.294} & \textbf{-103.668} & \textbf{101} & 981911 \\
   \bottomrule
   \end{tabular}
   
       }
       \label{tab:toygpu}
   \end{table}

   %%%%%%%%%%%%%%%%%%%%%%%%%%%%%%%%%%%%%%%%%%%%%%%%%%%%%%%%%%%%

   \newpage

   \section{License}
   \label{sec:license}
   The code and propose dataset will be publicly accessible. We include the following licenses for the raw data we used in this paper.
   
    \begin{itemize}
           \item CAN-Bus : \href{https://github.com/Tommydag/CAN-Bus-Controller/blob/master/LICENSE}{MIT}
           %\item iot shield :\href{}{}
           \item FPGA-CAN :\href{https://github.com/WangXuan95/FPGA-CAN/blob/main/LICENSE}{GPL-3.0}
           %\item subrisc:\href{}{}
           \item sha256:\href{https://github.com/secworks/sha256/blob/master/LICENSE}{BSD-2-Clause}
           %\item 8051:\href{}{}
           \item DE2\_CCD\_edge:\href{https://github.com/AngeloJacobo/FPGA_OV7670_Camera_Interface/blob/main/LICENSE}{MIT}
           %\item image porcessing:\href{}{}
           \item picorv:\href{https://github.com/YosysHQ/picorv32/blob/main/COPYING}{ISC}
           \item serv:\href{https://github.com/olofk/serv/blob/main/LICENSE}{ISC}
           \item mor1kx:\href{https://github.com/openrisc/mor1kx?tab=License-1-ov-file#readme}{CERN-OHL-W}
           %\item toygpu:\href{}{}
           \item ariane133:\href{https://github.com/The-OpenROAD-Project/OpenROAD-flow-scripts/blob/master/flow/designs/src/ariane136/LICENSE}{SOLDERPAD HARDWARE}
           \item ariane136:\href{https://github.com/The-OpenROAD-Project/OpenROAD-flow-scripts/blob/master/flow/designs/src/ariane136/LICENSE}{SOLDERPAD HARDWARE}
           \item bp:\href{https://github.com/The-OpenROAD-Project/OpenROAD-flow-scripts/blob/master/flow/designs/src/black_parrot/LICENSE}{BSD-3-Clause}
           \item bp\_be:\href{https://github.com/The-OpenROAD-Project/OpenROAD-flow-scripts/blob/master/flow/designs/src/bp_be_top/LICENSE}{BSD-3-Clause}
           \item bp\_fe:\href{https://github.com/The-OpenROAD-Project/OpenROAD-flow-scripts/blob/master/flow/designs/src/bp_fe_top/LICENSE}{BSD-3-Clause}
           \item swerv\_wrapper:\href{https://github.com/The-OpenROAD-Project/OpenROAD-flow-scripts/blob/master/flow/designs/src/swerv/LICENSE}{Apache }
           %\item chameleon:\href{}{}
                    
       \end{itemize}
   \end{document}